\documentclass[aip, amsmath,amssymb, reprint]{revtex4-1}

\usepackage{graphicx}
\usepackage{dcolumn}
\usepackage{bm}

\usepackage[utf8]{inputenc}
\usepackage[T1]{fontenc}
\usepackage{mathptmx}
\usepackage{mhchem}
\usepackage{notes2bib}

\begin{document}

\title {Understanding real-time time-dependent density-functional theory simulations of ultrafast laser-induced dynamics in organic molecules}

\author{Jannis Krumland}
\affiliation{Humboldt-Universit\"at zu Berlin, Physics Department and IRIS Adlershof, 12489 Berlin, Germany}
\author{Ana M. Valencia}
\affiliation{Humboldt-Universit\"at zu Berlin, Physics Department and IRIS Adlershof, 12489 Berlin, Germany}
\author{Stefano Pittalis}
\affiliation{CNR Nano, Istituto Nanoscienze, Centro S3, 41125 Modena, Italy}
\author{Carlo A. Rozzi}
\affiliation{CNR Nano, Istituto Nanoscienze, Centro S3, 41125 Modena, Italy}
\author{Caterina Cocchi}
 \affiliation{Humboldt-Universit\"at zu Berlin, Physics Department and IRIS Adlershof, 12489 Berlin, Germany}
 \affiliation{Carl von Ossietzky Universität Oldenburg, Institute of Physics, 26129 Oldenburg, Germany}
 \email{caterina.cocchi@uni-oldenburg.de}

\date{\today}
\begin{abstract}
Real-time time-dependent density functional theory, in conjunction with the Ehrenfest molecular dynamics scheme, is becoming a popular methodology to investigate ultrafast phenomena on the nanoscale.
Thanks to recent developments, it is also possible to explicitly include in the simulations a time-dependent laser pulse, thereby accessing the transient excitation regime.
However, the complexity entailed in these calculations calls for in-depth analysis of the accessible and yet approximate (either ``dressed'' or ``bare'') quantities, in order to evaluate their ability to provide us with a realistic picture of the simulated processes.
In this work, we analyze the ultrafast dynamics of three small molecules (ethylene, benzene, and thiophene) excited by a resonant laser pulse in the framework of the adiabatic local-density approximation. 
The electronic response to the laser perturbation in terms of induced dipole moment and excited-state population is compared to the results given by an exactly solvable two-level model. 
In this way, we can interpret the charge-carrier dynamics in terms of simple estimators,
such as the number of excited electrons.
From the computed transient absorption spectra we unravel the appearance of nonlinear effects such as excited-state absorption and vibronic coupling. 
In this way, we observe that the laser excitation affects the vibrational spectrum by enhancing the anharmonicities therein while the coherent vibrational motion contributes to stabilize the electronic excitation already within a few tens of femtoseconds. 
\end{abstract}

\maketitle
\section{Introduction}
Real-time time-dependent density-functional theory (RT-TDDFT) is a mature methodology to study atomistically the ultrafast dynamics of systems up to thousands of atoms.\citealp{andr+jpcm2012,10000atoms} 
Being intrinsically non-perturbative, this approach can access on the same footing both the linear and the nonlinear regime of excitations.\citealp{yaba+prb1996,cocc+prl2014}
Recent developments of RT-TDDFT have added the possibility to explicitly include a time-dependent electric field, \citealp{giov+2013chpch,lopa+jctc2011,yama+2019prb,sato+2014prb,jiao+2013pla,otob+2008prb,lee+2014jap,wa+2016epjb,wach+2014prl,matheusPaper} paving the way for RT-TDDFT to simulate pump-probe experiments and thus to investigate quantitatively the dynamics of charge carriers on their natural femtosecond timescale. 
Merging RT-TDDFT with the Ehrenfest molecular dynamics scheme \citealp{marq+ARPC2004,marq+2012,andr+2009jctc,rozz+jcpm2017,matheusPaper} additionally allows one to describe the coupled electron-nuclear dynamics in the sub-picosecond time window, adopting a mixed quantum-classical treatment. 

With the capabilities illustrated above, it is possible for RT-TDDFT to provide us with insights about excitation processes in the femtosecond regime which can be revealed nowadays by the most recent laser technologies. \citealp{ga+arpc2012,pazo+rmp2015,land+pr2015,somm+16nature,desi+16natcom,schl+18natp,buad+18}
The possibility to access the (transient) response of the system before, during, and after laser illumination, as well as the flexibility to decouple the effects of electronic and nuclear response, make the RT-TDDFT simulations complementary to laboratory observations. 

The interpretation of RT-TDDFT results, however, can be far from trivial.
First, the single-particle orbitals used to reproduce the dynamics of the particle density do not have any stringent physical meaning unless the effects of the electron-electron interactions are somewhat small.
Second, applications must unavoidably make use of approximations. 
Especially in dealing with large systems, restrictions to functional forms that can only capture local correlations, either in space, in time, or both, are often invoked to make calculations doable.
Consequently, additional analysis is needed to evaluate the validity of the corresponding results. 
Such a task is not simplified by the fact that accurate numerical benchmarks may not be available and that reference experiments often deal with novel phenomena that are non-trivial to interpret. 

In this work, we show that some additional and yet minimalist modeling
can help understanding and thus interpreting RT-TDDFT simulations.  
In particular, we investigate the laser-induced dynamics of three small molecules (ethylene, benzene, and thiophene) which can be viewed as the building blocks of larger organic complexes of experimental and technological interest. 
For this purpose, we employ RT-TDDFT in the adiabatic local-density approximation (ALDA),\citealp{zang-sove80pra,ekar1984prb,ekar1984prl,perd-wang92prb} which is widely used especially in calculations of large systems due its reduced computational costs compared to more advanced approximations.
After the analysis of the linear absorption properties of the three molecules, we investigate their response to a resonant femtosecond (fs) laser pulse, monitoring relevant quantities such as the induced dipole moment and photo-excited charge-carrier population. 
The comparison with the results obtained in the same conditions from an exactly solvable two-level model offers a useful diagnostic tool to interpret the laser-induced electron dynamics from RT-TDDFT and to highlight the physical meaning of the corresponding outcomes. 
The analysis of the transient absorption spectra allows us to illustrate the capability of RT-ALDA to reproduce entailed nonlinear processes such as excited-state absorption. 
Finally, the effects of the coupled electron-nuclear dynamics are discussed in details to disclose the role of coherent nuclear motion and vibronic coupling in stabilizing the excitation in the fs time window. 
 
This paper is organized as follows. 
In Section~\ref{sec:methods} we review the methodology and report the computational details. 
Section~\ref{sec:results} contains the body of the results, starting from the linear absorption properties of the systems (Section~\ref{subsec:linear}) and proceeding with the outcomes of the laser-induced dynamics of the electrons (Section~\ref{dyn_sim.sec}), the transient absorption spectra (Section~\ref{tas.sec}), excited-state absorption (Section~\ref{sec:esa}), and vibronic coupling (Section~\ref{vibr.sec}). 
The conclusions are reported in Section~\ref{sec:conclu}.
\section{Methodology}
\label{sec:methods}
\subsection{Theoretical background}
The results presented in this work are obtained in the framework of time-dependent DFT,\citealp{rung+1984prl} adopting its real-time implementation to access the electron dynamics.
Within RT-TDDFT, the time-dependent (TD) electron density for a spin-unpolarized system
\begin{equation}
\rho(\textbf{r},t) = 2\sum\limits_i^{\text{occ}}|\phi_i(\textbf{r},t)|^2
\label{eq:rho}
\end{equation}
is calculated through the propagation of the Kohn-Sham orbitals $\phi_i(\textbf{r},t)$ according to the TD Kohn-Sham equation
\begin{equation}
i\frac{\partial}{\partial t}\phi_i(\textbf{r},t) =\left( -\frac{\nabla^2}{2}+v_{\text{KS}}[\rho](\textbf{r},t)\right)\phi_i(\textbf{r},t),
\end{equation}
where
\begin{equation}
\label{kspot.eq}
v_{\text{KS}}[\rho](\textbf{r},t) = v_{\text{en}}(\textbf{r},t) + v_{\text{ext}}(\textbf{r},t) + \int\text d^3r'\frac{\rho(\textbf{r}',t)}{|\textbf{r}-\textbf{r}'|} + v_{\text{xc}}[\rho](\textbf{r},t)
\end{equation}
is the effective potential. 
The first two terms describe the interaction of the electrons with the nuclei and with external fields, respectively, while the last two account for the electron-electron interactions. 
The exchange-correlation (XC) potential, $v_{xc}$, must be approximated, as its exact form is not known. 
This term is nonlocal in space and time,\citealp{mait+2002prl} meaning that it depends not only on the instantaneous density, but also on the density at earlier times.  

While efforts have been directed towards taking into account memory effects,\citealp{mait+2002prl} in most calculations, the adiabatic approximation is still assumed, \textit{i.e.}, the XC potential is approximated using a ground-state (GS) XC potential with the time-dependent (TD) density. 
Here, we adopt the ALDA, where $v_{xc}$ is assumed to be local in space and time. 
The locality in space, which is inherited from the local-density approximation for $v_{xc}$ in the GS calculation, is known to impair the ability of TDDFT to correctly predict the energy of charge-transfer excitations\citealp{grit+jcp2004,dreu+jacs2004,autscpc2009,stei+jaft2009,rohr+jcp2009,mait2017jpcm,camp+jcc2017} and excitonic features.\citealp{dreu+2003jcp,tozejcp2003,maitjcp2005,cocc-drax15prb} 
The locality in time stems from the adiabatic approximation, \textit{i.e.}, the approximation of the TD XC functional as a GS XC functional of the instantaneous density. 
This assumption leads to a neglect of memory effects, which are thought to become crucial especially in real-time propagation schemes, where the system can be driven far away from its GS.\citealp{mait+2002prl, elli+chemphys2011,fuks+2011prb,fuks+pccp2014}
For these reasons, the accuracy of the ALDA results is often questioned (see, for example, Refs.~\citenum{ullrich+prb2006} and \citenum{bostr2018nano}).
Nonetheless, the ALDA remains a popular approach especially thanks to its numerical simplicity, which enables the treatment of systems otherwise out of reach by means of more accurate methods.

The coupling of the electronic system to an external TD electric field, $\textbf{E}(t)$, is described semi-classically in the dipole approximation and within the length gauge, such that the second term in Eq. \eqref{kspot.eq} is given by $v_{\text{ext}}(\textbf{r},t) = \textbf{r}\cdot\textbf{E}(t)$.
Absorption spectra are computed according to the scheme proposed in Ref.~\citenum{yaba+prb1996}, \textit{i.e.}, the system is probed by an instantaneous, broadband electric field, commonly referred to as a ``kick'':  
\begin{equation}
\textbf{E}(t)=\textbf{E}_{\text{probe}}(t)=\hat{\textbf{n}}\kappa\delta(t),
\end{equation}
where $\textbf{n}$ defines the polarization direction, $\kappa$ the kick strength, and $\delta(t)$ is the Dirac delta function. 
Perturbing the system with a kick is equivalent to multiplying each occupied  orbital with a phase factor:\citealp{yaba+prb1996}
\begin{equation}
\phi_i(\textbf{r},0^+) = e^{i\kappa\hat{\textbf{n}}\cdot\textbf{r}}\phi_i(\textbf{r},0^-).
\end{equation}
To calculate the linear absorption spectrum, we start from the GS  wave functions, $\phi_i(\textbf{r},0^-)=\phi_i(\textbf{r})$.
After the instantaneous perturbation given by the kick, the system is propagated for a certain amount of time. 
The total duration of the propagation determines the achieved spectral resolution. 
The resulting induced dipole moment,
\begin{equation}\label{dipole.eq}
\boldsymbol{d}(t) = 
-\int\text d^3r\, \textbf{r}\left[\rho(\textbf{r},t) - \rho_g(\textbf{r})\right],
\end{equation}
is a superposition of free oscillations with frequencies and amplitudes corresponding to the energies and oscillator strengths of the excitations of the system, respectively. 
Here, $\rho(\textbf{r},t)$, defined in Eq.~\eqref{eq:rho}, and $\rho_g(\textbf{r})$ are the TD and GS electron densities, respectively.
The absorption spectrum is obtained by taking the imaginary part of the Fourier transform of $\boldsymbol{d}(t)$ as obtained from Eq. \eqref{dipole.eq}. 
If $\kappa$ is sufficiently small, only the linear response is obtained, while stronger perturbations excite in higher order.\citealp{cocc+prl2014, albertosPaper}

In order to calculate the response of the material to an ultrafast laser pulse, we consider electric fields of the form
\begin{equation}\label{pulse.eq}
\textbf{E}(t)=\textbf{E}_{\text{pump}}(t) = \hat{\textbf{n}}E_0f(t)\cos(\omega_pt).
\end{equation}
Here, $E_0$ is the peak amplitude, $f(t)$ an envelope function, and $\omega_p$ the carrier frequency, which can be tuned to be in resonance with specific excitation energies of the system. 
It is possible to combine this realistically shaped laser pulse and the instantaneous kick mentioned above in order to simulate pump-probe experiments.\citealp{giov+2013chpch}
In this case, the electric field can be written as
\begin{equation}\label{pumpprobe.eq}
\textbf{E}(t) = \textbf{E}_{\text{pump}}(t) + \textbf{E}_{\text{probe}}(t-t_0),
\end{equation}
where the first term, referred to as the ``pump'', prepares the system in a non-stationary state, and the second one, the ``probe'', interrogates it at a later time $t_0$. 
In this way, it is possible to gain information about excited and transient states of the system. 

The dipole moment induced by the two fields includes interference effects depending on their phase relation. These effects are not measured in typical pump-probe experiments, as they vanish when taking the average over the response of the system to multiple laser shots.\citealp{seid+1995jcp}
To obtain a transition dipole moment that is insensitive to the aforementioned phase difference, we adopt the so-called ``phase-cycling approach''\citealp{hamm_zanni_2011} which consists in performing four simulations applying the field in Eq. \eqref{pumpprobe.eq}, adding each time a constant phase of $\pi/2$ to the pump field.
In this way, the applied laser field becomes:
\begin{equation}
    \textbf{E}_{\text{pump}}^{(j)}(t)=\hat{\textbf{n}}E_0f(t)\cos(\omega_pt+j\pi/2),
\end{equation}
where $j$ runs from 1 to 4.
The four resulting total dipole moments are subsequently averaged and the pump-only contributions are subtracted. 
A sine transformation finally yields the absorption spectrum. 
In order to calculate the transient absorption spectrum (TAS), this procedure is repeated for different probe delay times $t_0$, corresponding to instants before, during, and after the application of the pump.

The non-perturbative approach provided by RT-TDDFT does not offer any information about the single-particle transitions that contribute to the absorption in the linear regime. 
For this reason, this method is often complemented by the solution of the TD Kohn-Sham equations in the linear-response scheme first proposed by Mark Casida.\citealp{casi1996rda}
Within this framework, the problem is linearized and reformulated in terms of the matrix equation\citealp{casi+2012arpc}
\begin{equation}\label{casida.eq}
 \begin{pmatrix}
 \textbf{A} & \textbf{B} \\ \textbf{B}^* & \textbf{A}^*
 \end{pmatrix}
 \begin{pmatrix}
 \vec{X} \\ \vec{Y} 
 \end{pmatrix}
 = \omega
  \begin{pmatrix}
  \textbf{1} & \textbf{0} \\ \textbf{0} & \textbf{-1}
  \end{pmatrix}
   \begin{pmatrix}
   \vec{X} \\ \vec{Y} 
   \end{pmatrix}.
\end{equation}
Here, $\vec{X}$ and $\vec{Y}$ are vectors including the excitation and de-excitation coefficients $X_{im}$ and $Y_{im}$, respectively, which correspond to transitions from the occupied  orbitals $\phi_i(\textbf{r})$ to the unoccupied ones $\phi_m(\textbf{r})$, and vice versa. 
The eigenvalue $\omega$ represents the excitation energy, and, within the adiabatic approximation, the matrices $\textbf{A}$ and $\textbf{B}$ are given by
\begin{equation}
A_{im,jn} = \delta_{i,j}\delta_{m,n}(\varepsilon_m-\varepsilon_i)+\langle im | f_{\text{Hxc}}| nj \rangle 
\label{eq:A}
\end{equation}
and 
\begin{equation}
B_{im,jn} = \langle im | f_{\text{Hxc}}| jn \rangle,
\label{eq:B}
\end{equation}
respectively. 
In Eq.~\eqref{eq:A}, $\varepsilon_k$ are the KS eigenvalues, and the Hartree-XC kernel appearing in Eqs.~\eqref{eq:A} and ~\eqref{eq:B} reads:
\begin{equation}
f_{\text{Hxc}}(\textbf{r},\textbf{r}') = \frac{1}{|\textbf{r}-\textbf{r}'|} + \frac{\delta v_{\text{xc}}[\rho](\textbf{r})}{\delta\rho(\textbf{r}')}\bigg|_{\rho=\rho_g},
\end{equation}
where $v_{\text{xc}}[n](\textbf{r})$ is the GS XC potential and $\rho_g(\textbf{r})$ the GS electron density [already introduced in Eq.~\eqref{dipole.eq}].
Eq.~\eqref{casida.eq} yields the excitation energies $\omega_e$ and the corresponding transition weights $\vec{X}^{(e)}$ and $\vec{Y}^{(e)}$ of the excited many-body states $|e\rangle$. 

Given the known limitations of ALDA in describing excitonic effects, \citealp{cocc-drax15prb, dreu+2003jcp,tozejcp2003,maitjcp2005} we employ many-body perturbation theory (MBPT) as a benchmark for the linear absorption spectra. 
To this end, we solve the Bethe-Salpeter equation (BSE) for a system of quasiparticles with energies calculated within the  $G_0W_0$ approximation.\citealp{hedi+1965pr,fabe+13jcp}
Formally, the BSE is the equation of motion for the two-particle polarizability, $L = L_0 + L_0KL$, \citealp{onid+02rmp} where the interaction kernel $K$ includes the repulsive exchange term given by the bare Coulomb potential and the statically screened electron-hole Coulomb attraction. 
For further details about this formalism, see \textit{e.g.}, Refs. \citenum{brun+16cpc,fabe+13jcp,cocc-drax15prb,brun+jcp2015}.

One of the advantages of RT-TDDFT is the possibility to be interfaced with molecular dynamics, \textit{e.g.}, adopting the Ehrenfest scheme.\citealp{marq+ARPC2004,andr+2009jctc,marq+2012} The Ehrenfest formalism is a mean-field approach, with the forces on the classical trajectory being calculated by averaging over the quantum-mechanical degrees of freedom.\citealp{rozz+jcpm2017}
The (classical) equation of motion for the $J$th nucleus is herein given by\citealp{ullrich}
\begin{equation}
M_J\frac{\text d^2}{\text dt^2}\mathbf{R}_J = -\nabla_{\textbf{R}_J}\left[v_{\text{ext}}(\textbf{R}_J,t)+v_{nn}(R(t))+\int\text d^3r\,\rho(\textbf{r},t)v_{en}(\textbf{r},R(t))\right],
\end{equation}
where $M_K$ and $\textbf{R}_K$ are the mass and position of the $K$th nucleus, respectively, $R(t)$ stands for the whole set of nuclear coordinates, and $v_{nn}(R(t))$ is the electrostatic energy between the nuclei.

\subsection{Computational details}
All RT-TDDFT calculations in this work are performed with the \textsc{octopus} code.\citealp{marq+cpc2003,cast+pssb2006,andr+pccp2015,Octopus2020}
Core electrons are approximated by norm-conserving Troullier-Martins pseudopotentials,\citealp{trou-mart91prb} calculated within the LDA in the Perdew-Zunger parameterization.\citealp{perd-zunger81prb}
The pseudo-orbitals (hereafter referred to as \textit{orbitals} for brevity) are represented on a real-space grid with a spacing of 0.3~bohr inside a simulation box built by interlocking spheres with a radius of 12~bohr centered at each atom. 
The orbitals are propagated using the approximated enforced time-reversal symmetry scheme\citealp{cast+jcp2004} with a time step of 1.2~attoseconds. 
To obtain the absorption spectra, each system is propagated for 10~fs after the application of the kick, resulting in a full width at half maximum of the peaks of 435~meV. 
For the laser fields, we use a peak amplitude $E_0$ corresponding to an intensity of $5\times 10^{11}$~W/cm$^2$ (unless stated otherwise), and a Gaussian envelope function $f(t)$ with standard deviation of 2~fs. 
The carrier frequency, $\omega_p$, is chosen in resonance with specific excitations in each system, identified in their RT-ALDA spectra. 
TAS are calculated by probing the system at intervals of 1~fs. 
In the RT-ALDA calculations including coupled electron-nuclear dynamics, we start from a geometry relaxed using the FIRE algorithm\citealp{bitz+prl2006} minimizing the forces until they are lower than $10^{-3}$~Ha/bohr. 
No initial velocities are assigned to the nuclei unless stated otherwise.

Linear-response spectra are computed with the MOLGW code\citealp{brun+16cpc}, using Gaussian-type cc-pVQZ basis sets.\citealp{brun12jcp}
To calculate the MBPT spectra, the resolution-of-identity approximation\citealp{weig+02} is employed. 
The range-separated hybrid functional CAM-B3LYP\citealp{yana+2004cpl} is used in the underlying DFT calculation as an improved starting point for the single-shot $G_0W_0$ approximation 
.\citealp{fabe+13jcp,brun+jctc2013}

\section{Results and Discussion}
\label{sec:results}
\subsection{Linear absorption spectra}
\label{subsec:linear}
\begin{figure}
\includegraphics[width=.5\textwidth]{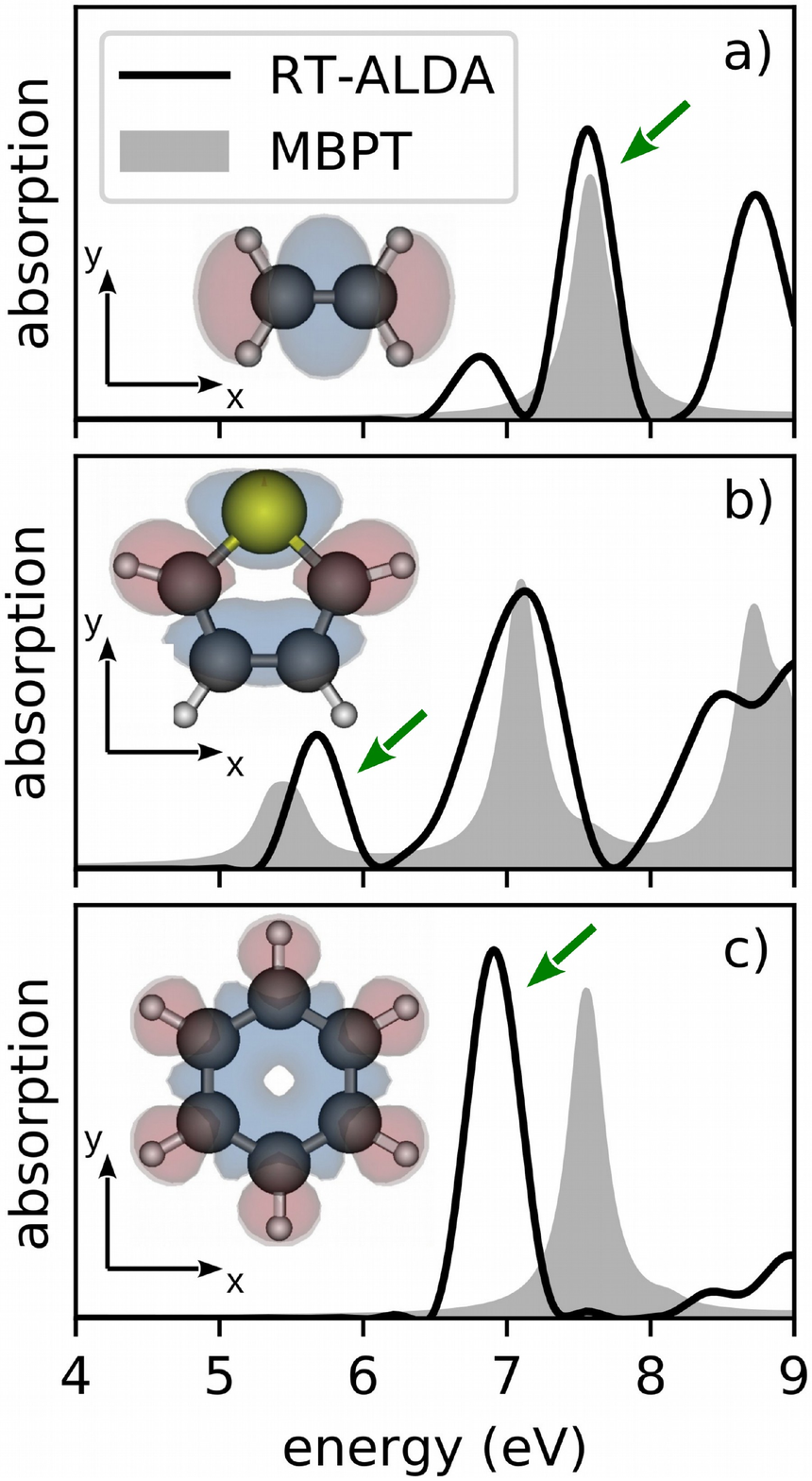}
\caption{Absorption spectra of a) ethylene, b) thiophene, and c) benzene calculated from MBPT and RT-ALDA. The excitations targeted by the resonant pulse are highlighted by green arrows. Insets: isosurfaces of the difference density ($\Delta\rho^{(e)}(\textbf{r})$ ranges from $1\times 10^{-3}$~\AA$^{-3}$ to $5\times 10^{-3}$~\AA$^{-3}$ depending on the molecule) related to the these excitations on top of ball-and-stick representations of the molecules (black, white, and yellow spheres stand for carbon, hydrogen, and sulfur atoms, respectively). Blue and red colors in the isosurfaces denote electron depletion and accumulation, respectively.}
\label{spectra.fig}
\end{figure}

We start our analysis by examining the linear absorption spectra of ethylene, thiophene, and benzene, depicted in Fig.~\ref{spectra.fig}, as obtained from RT-ALDA and MBPT. 
Focusing on the pronounced maximum at around 7.6 eV in the spectrum of ethylene [Fig. \ref{spectra.fig}a)], we find a good match between the RT-ALDA and the MBPT results, and excellent agreement with the experimentally determined value of 7.63 eV.~\cite{ethSpectrum} 
This peak is given by a $\pi\rightarrow\pi^*$ excitation and has transition dipole moment (TDM) in the $x$ direction [see coordinate systems in the inset of Fig. \ref{spectra.fig}a)]. 
According to the linear-response calculations (see Supporting Information, Table S1 and Fig. S2), this excitation corresponds mainly to a transition from the highest occupied molecular orbital (HOMO) to the lowest unoccupied molecular orbital (LUMO). 
To visualize the character of this excitation, we plot the corresponding \textit{difference density},\citealp{dreu+cr2005} calculated as 
\begin{equation}
\Delta\rho^{(e)}(\textbf{r}) = \sum_{m}^{\text{unocc}}\sum_{i}^{\text{occ}}|X_{im}^{(e)}|^2\left(|\phi_m(\textbf{r})|^2 - |\phi_i(\textbf{r})|^2\right),
\label{static_ind_dens.eq}
\end{equation}
where the transition weights $X_{im}^{(e)}$ are obtained from the solution of the Casida's equation, Eq. \eqref{casida.eq}. 
On the left-hand-side, $\Delta\rho^{(e)}(\textbf{r})$ represents the difference between the excited-state and the ground-state electron densities. 
In the case of ethylene, we see that the electron density accumulates on the opposite ends of the molecule along the direction of the carbon double bond [inset of Fig. \ref{spectra.fig}a)]. 
This characteristic is particularly important in the analysis of the coupled electron-nuclear dynamics, which is discussed in Section \ref{vibr.sec} below. 
The weak peak at lowest energy in the RT-ALDA spectrum [Fig. \ref{spectra.fig}a)] is given by the transition from the HOMO to the LUMO+1, and has a TDM oriented in the $z$ direction. 
Due to the small overlap between the involved orbitals (see Fig. S1), the excitation energy is underestimated by ALDA, which notoriously fails to correctly predict the energy of charge-transfer excitations. \citealp{dreu+2003jcp,dreu+jacs2004,autscpc2009,camp+jcc2017,maitjcp2005} 
Indeed, in the MBPT spectrum, the excitation appears above the absorption onset, as a shoulder of the main peak around 8~eV. Moreover, we have checked that applying  a self-interaction correction\citealp{Legrand2002} to ALDA alleviates this issue (see Supporting Information, Fig.~S7).
These two excitations have TDM perpendicular to each other, such that it is possible to selectively target either excitation with an external electric field. Hence, the incorrect order of the first two excitations will not affect the following analysis of laser-induced electron dynamics (see Sec. S4 and Figs. S8 and S9). Due to the very small overlap between the LUMO and LUMO+1, interactions are likely negligible also if the nuclear motion is taken into account. The higher-energy peak at around 8.7~eV in the ALDA spectrum is found in the MBPT spectrum at a higher energy, outside the considered window. 

In the case of thiophene [Fig. \ref{spectra.fig}b)], the agreement between the MBPT and TDDFT spectra is good, especially for the first peaks.
The first maximum at about 5.6~eV comprises two bright excitations energetically very close to each other. 
The lower energy one has a TDM oriented in the $y$ direction, while the second one is polarized along $x$. 
The analysis of the linear-response calculation (see Table S2) shows that the first excitation corresponds mainly to the transition from the HOMO-1 to the LUMO, but has a significant contribution from the HOMO to the LUMO+3 as well. 
The $x$-polarized excitation is given mostly by the HOMO-LUMO transition. The MBPT value for the energy of the first peak (5.37~eV) comes closer to the experimental one in solution (5.34~eV).~\cite{thioSpectrum}

Considering now benzene [see Fig. \ref{spectra.fig}c)], we see that the RT-ALDA and MBPT spectra match qualitatively. 
Both methods predict an intense peak formed by a doubly degenerate excitation with contributions from transitions between the doubly degenerate HOMO and LUMO (see Table S3). 
This excitation is again a $\pi\rightarrow\pi^*$ transition with the electron density accumulating on the outer part of the aromatic ring, as shown in the inset of Fig. \ref{spectra.fig}c). In this case, RT-ALDA actually reproduces the experimental value for the first peak at 6.94~eV~\cite{benzeneSpectrum} more accurately than MBPT.
\subsection{Laser-induced dynamics}
\label{dyn_sim.sec}
\begin{figure}
	\includegraphics[width=.45\textwidth]{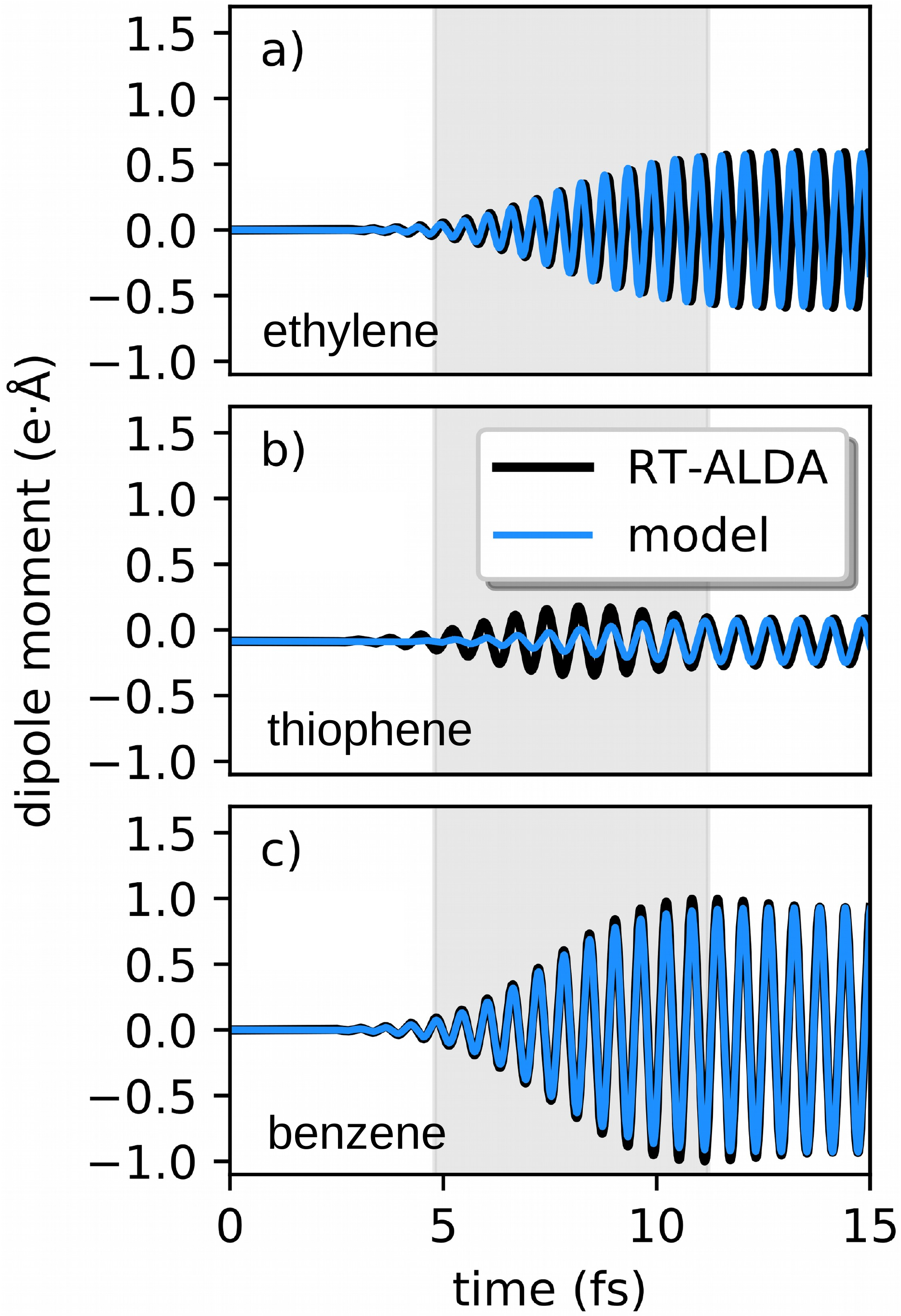}
	\caption{Induced dipole moments [Eq. \eqref{dipole.eq}] of a) ethylene, b) thiophene, and c) benzene upon laser illumination in the time window indicated by the grey shaded areas. The frequency and polarization of the respective pulses are tuned to the excitations highlighted by green arrows in Fig.~\ref{spectra.fig}.}
	\label{dipole.fig}
\end{figure}

With the knowledge of the linear absorption spectra, in the next step we investigate the electron dynamics induced in the molecules by an ultrafast laser pulse in resonance with the excitations highlighted in Fig. \ref{spectra.fig} and discussed above. 
In particular, we analyze the induced dipole moment and the excited-state population.
To interpret our findings, we use as a point of comparison an exactly solvable problem, consisting of a two-level system perturbed to the same laser pulse. 
While this model is by itself subject to the crude approximation of two-level excitations, its solution is exact and its physical meaning transparent.

The Hamiltonian of the two-level system is given by
\begin{equation}\label{model.eq}
\hat H(t) = E_g|g\rangle\langle g|+E_e|e\rangle\langle e| - \textbf{d}_{ge}\cdot\textbf{E}(t)\big(|g\rangle\langle e|+|e\rangle\langle g|\big),
\end{equation}
where $|g\rangle$ and $|e\rangle$ are the GS and excited state (ES), respectively; the TDM $\textbf{d}_{ge}$ is taken from the Casida calculation and the excitation energy $(E_e-E_g)$ stems from the RT-ALDA spectrum. 
We neglect the terms related to the permanent dipole of the molecules in the GS ($\textbf{d}_g$) and in the ES ($\textbf{d}_e$), which would contribute the additional terms $-\textbf{d}_g\cdot\textbf{E}(t)|g\rangle\langle g|$ and $-\textbf{d}_e\cdot\textbf{E}(t)|e\rangle\langle e|$, since they do not have a significant impact upon the time evolution. 
The coefficients $c_g(t)$ and $c_e(t)$ in the TD wavefunction,
\begin{equation}
|\Psi(t)\rangle = c_g(t)e^{-iE_gt}|g\rangle + c_e(t)e^{-iE_et}|e\rangle,\label{superposition.eq}
\end{equation}
are calculated numerically using the Euler method. 
These coefficients remain constant once the field is turned off. 
To relate this model to the results provided by RT-ALDA, we analyze the two-level system in terms of the electron density, which is obtained from the many-body wavefunction $\Psi(\textbf{r}_1,\hdots,\textbf{r}_N,t) = \langle \textbf{r}_1\hdots \textbf{r}_N|\Psi(t)\rangle$ as
\begin{equation}
\rho(\textbf{r},t) = N\int\text d^3r_2\hdots\text d^3r_N\,|\Psi(\textbf{r}, \textbf{r}_2,\hdots,\textbf{r}_N,t)|^2.
\label{wf_to_rho.eq}
\end{equation}
Inserting the wavefunction in the form of Eq. \eqref{superposition.eq} into Eq. \eqref{wf_to_rho.eq}, we find 
\begin{align}
\delta\rho(\textbf{r},t) &= \rho(\textbf{r},t) - \rho_g(\textbf{r})\nonumber\\
&= |c_e(t)|^2\Delta\rho^{(e)}(\textbf{r})
+2\Re\left[ c_g^*(t)c_e(t)e^{-i(E_e-E_g)t}\rho_t^{(e)}(\textbf{r})\right],\label{delta_rho_model.eq}
\end{align}
where the the transition density is defined following Ref. \citenum{dreu+cr2005}:
\begin{align}\label{transition_density.eq}
\rho_t^{(e)}(\textbf{r}) = N\int\text d^3r_2\hdots&\text d^3r_N\,\Psi_g^*(\textbf{r},\textbf{r}_2,\hdots,\textbf{r}_N)\Psi_e(\textbf{r},\textbf{r}_2,\hdots,\textbf{r}_N).
\end{align}
The difference density $\Delta\rho^{(e)}(\textbf{r}) = \rho_e(\textbf{r}) - \rho_g(\textbf{r})$ corresponds to Eq. \eqref{static_ind_dens.eq}, if the ES is represented as a superposition of Slater determinants with coefficients obtained from the linear-response calculation. 
From the analytic solution of the two-level problem, it is known that the excited-state population $|c_e(t)|^2$ changes on a timescale determined by the \textit{intensity} of the field, whereas the coherence $c_g^*(t)c_e(t)e^{-i(E_e-E_g)t}$ oscillates with the \textit{frequency} of the resonant field. 
Hence, the induced density is the sum of two terms: the first one is related to $\Delta\rho^{(e)}(\textbf{r})$ and changing at a rate determined by the field strength, while the second one gives rise to the coherent electronic motion of a superposition state, with a spatial profile given by $\rho_t^{(e)}(\textbf{r})$ [Eq. \eqref{transition_density.eq}].
In analogy to Eq. \eqref{delta_rho_model.eq}, the induced dipole moment is given by
\begin{align}
\boldsymbol{d}(t) &= \langle\Psi(t)|\hat{\boldsymbol{d}}|\Psi(t)\rangle \nonumber\\&=|c_e(t)|^2 \left(\textbf{d}_e-\textbf{d}_g\right)+2\Re\left[c_g^*(t)c_e(t)e^{i(E_e-E_g)t}\textbf{d}_{ge}\right].
\label{dipole_model.eq}
\end{align}
The induced dipole moment obtained from the two-level model [Eq. \eqref{dipole_model.eq}] can be compared to that from the RT-ALDA calculation [Eq.~\eqref{dipole.eq}].

For ethylene the results obtained with the two methods agree well [see Fig. \ref{dipole.fig}a)]. 
In both cases, the dipole moment builds up during laser irradiation and remains constant afterwards. 
A careful inspection of Fig. \ref{dipole.fig}a) shows a frequency shift that increases over time and becomes clearly visible towards the end of the simulation. 
This behavior is likely due to a shift of the resonance caused by pumping the system (see also Sec. \ref{tas.sec}), which is an unphysical artifact of the adiabatic approximation.\citealp{fuks+prl2015,luo2016, prov+2015jctc} 
In the case of thiophene, the dipole moment computed from RT-ALDA has an additional contribution during the active time window of the laser [Fig. \ref{dipole.fig}b)]. 
It represents a non-dissipative, temporary change of the electronic density and is related to the real part of the polarizability, which in the case of thiophene is relatively large in comparison to its imaginary part. 
This effect is not included in the two-level model.
After the field is switched off, the results of the model and of RT-ALDA are in excellent agreement. 
In the case of benzene, the envelope of the dipole moment increases once the field is applied [Fig. \ref{dipole.fig}c)]. 
In the two-level model, the envelope stays constant afterwards, whereas in the RT-ALDA result a slight oscillatory modulation of the amplitude is observed. 
These quantum beats suggest that the evolution of the dipole moment for this system can involve a three-state superposition including the GS and two excited states, which must be close in energy since the beat frequency is low.

\begin{figure}
\includegraphics[width=.45\textwidth]{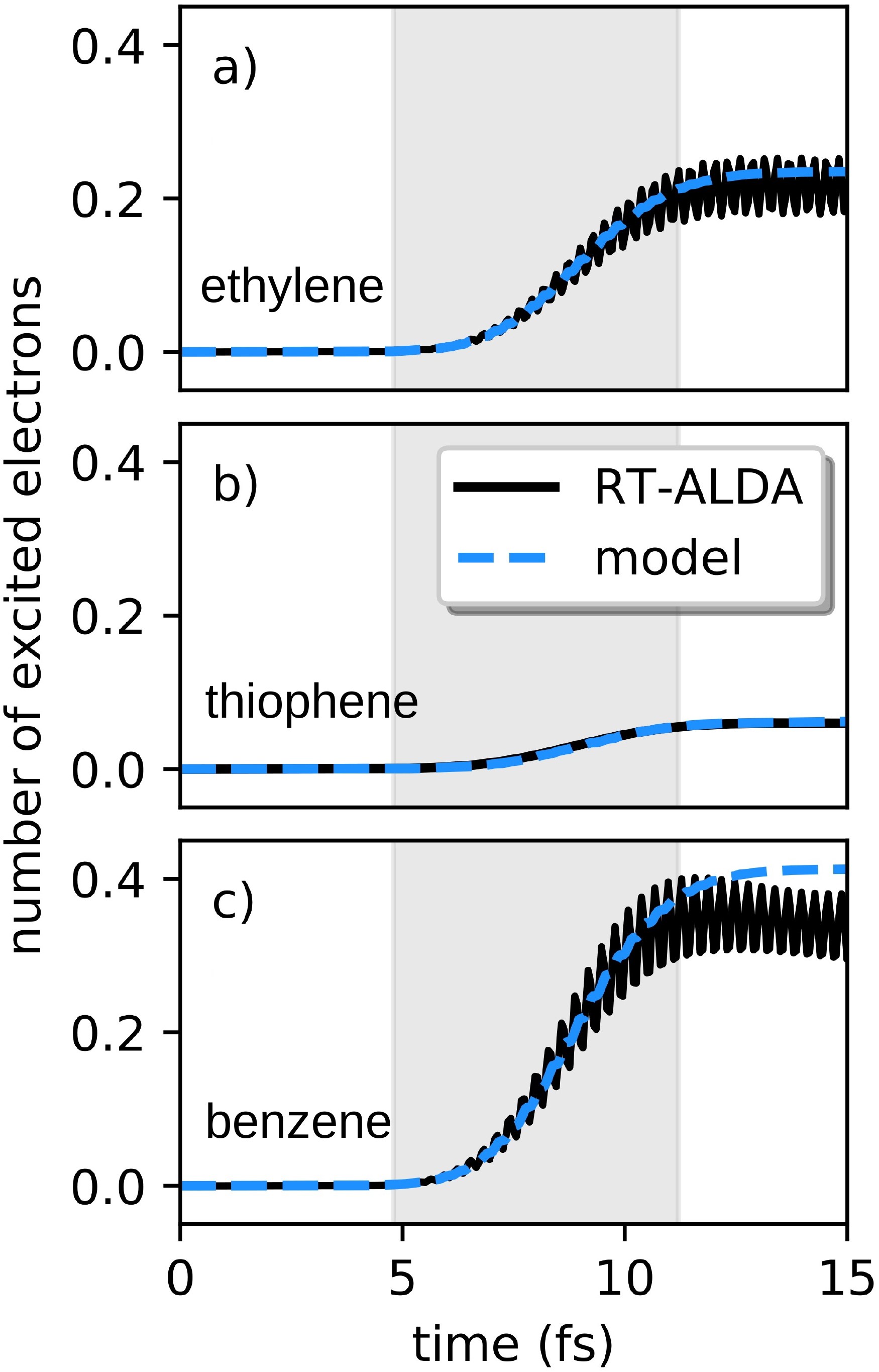}
	\caption{Number of excited electrons in a) ethylene, b) thiophene, and c) benzene calculated from the orbital projections given by RT-ALDA (dashed blue) [Eq. \eqref{pop.eq}] compared to the excited-state populations obtained from the solution of the two-level problem [black, see Eq. \eqref{model_pop.eq}]. The grey shaded areas indicate the active window of the laser pulse.}
	\label{population.fig}
\end{figure}

Next, we compare the ES population obtained from the model system,
\begin{equation}
P_e(t) = |c_e(t)|^2,
\label{model_pop.eq}
\end{equation}
with the \textit{number of excited electrons}, $N_{\text{ex}}(t)$, computed from RT-ALDA in an independent-particle manner, \textit{i.e.}, by projecting all occupied time-dependent orbitals onto the static unoccupied ones,\citealp{otob+2008prb}
\begin{align}\label{pop.eq}
N_{\text{ex}}(t) &= 2\sum\limits_m^{\text{unocc}}\sum\limits_j^{\text{occ}}\left|\langle\phi_m(0)|\phi_j(t)\rangle\right|^2\nonumber\\
&= N - 2\sum\limits_{i,j}^{\text{occ}}\left|\langle\phi_i(0)|\phi_j(t)\rangle\right|^2,
\end{align}
where $N$ is the total number of electrons and the normalization of the TD orbitals is used:
\begin{equation}
    \sum\limits_{i}^{\text{occ}}|\langle\phi_i(0)|\phi_j(t)\rangle|^2 + \sum\limits_{m}^{\text{unocc}}|\langle\phi_m(0)|\phi_j(t)\rangle|^2 = 1.
\end{equation}
In case only one single excitation is affected by the laser, $P_e(t)$ and $N_{\text{ex}}(t)$ should be comparable, as the transition from $|g\rangle$ to $|e\rangle$ is a single-electron excitation. 
The curves in Fig. \ref{population.fig} obtained from RT-ALDA generally agree well with the ones given by the model. 
For all three molecules, both RT-ALDA and the model feature an increasing number of excited electrons as the laser is applied, and remain nearly constant afterwards. 
An excellent agreement between the results obtained with the two methods is found for thiophene [Fig. \ref{population.fig}b)]. 
On the other hand, differences appear in the case of benzene [see Fig. \ref{population.fig}c)]: $N_{\text{ex}}(t)$ (RT-ALDA, black) and $P_e(t)$ (two-level model, dashed blue) differ after the pulse, with $N_{\text{ex}}(t)$ decreasing slightly, while $P_e(t)$ remains constant. 
This behavior of $N_{\text{ex}}(t)$ mirrors the decrease observed in the envelope of the induced dipole moment [Fig. \ref{dipole.fig}c)]. 

It is worth noting that the orbital populations from RT-ALDA continue to evolve in time after the laser pulse, exhibiting rapid oscillations. These are due to the Hartree and XC potentials [Eq. \eqref{kspot.eq}] still being time-dependent, even without the external field, since the system is left in a non-stationary state. The charge density sloshes with the excitation frequency and through the Hartree and XC potentials this motion creates a time-dependent field in which the orbitals evolve.
In turn, this causes the populations to oscillate at exactly twice this frequency and shows the limits of the interpretation of the DFT orbital populations as the populations of the (dressed) many-body states.

\begin{figure*}
\includegraphics[width=\textwidth]{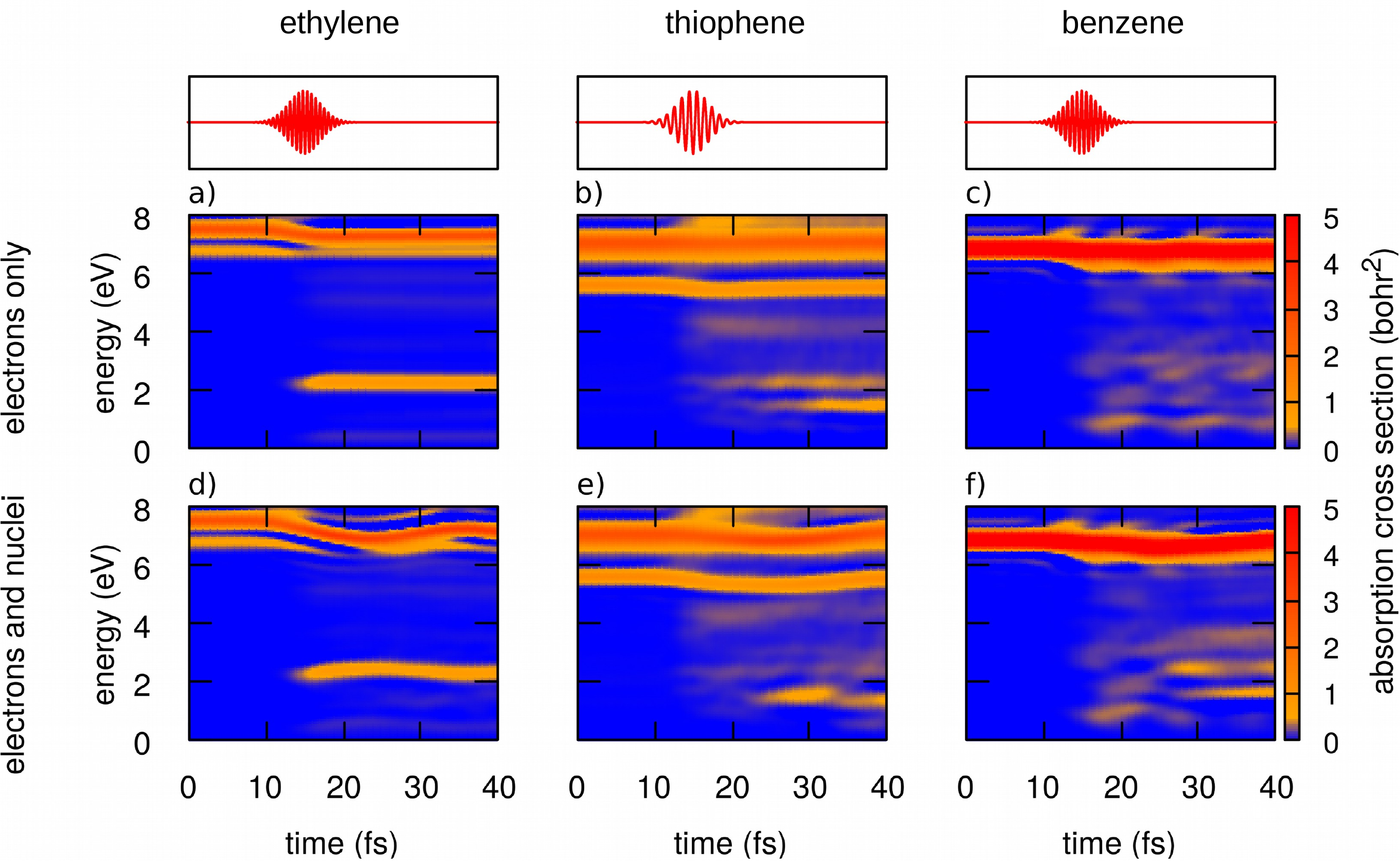}
\caption{Transient absorption spectra of ethylene (left), thiophene (center), and benzene (right). The upper panels (a-c) show the results when the nuclei are fixed, the lower panels (d-f) when the nuclei are moving according to the Ehrenfest dynamics. The color map is chosen to emphasize subtle features associated with a small magnitude of the absorption cross section.}
\label{tas.fig}
\end{figure*}

The observation that the number of excited electrons deduced directly from the bare KS quantities [Eq. \eqref{pop.eq}] generally compares well to the populations given by the interacting two-level system [Eq. \eqref{model_pop.eq}] contrasts with previous results, which report qualitatively incorrect population dynamics from adiabatic TDDFT.\cite{fuks+2011prb, fuks2013, fuks2014pra, fuks+pccp2014, ragh2011, ragh2012_2, mait2017jpcm,prov2016} We attribute the well-behaved nature of our simulations to meeting two conditions: i) avoiding charge-transfer excitations and ii) using a weak, finite laser pulse. Avoiding charge-transfer excitations is necessary, as it has been shown that charge-transfer dynamics are difficult to be captured by XC approximations which, as advanced as they can be, have been developed
for the ground state.\cite{elli+2012prl, fuks2013, mait2016, mait2017jpcm} Problems are less drastic in the case of local excitations, \textit{i.e.}, excitations exhibiting enhanced overlap between the orbitals of the initial and final state.
Still, Rabi oscillations cannot be correctly described for these excitations, either. Even if a continuous-wave laser is exactly in resonance with an excitation, the orbital populations exhibit only off-resonance Rabi oscillations (Fig. S9 and S10 in the Supporting Information). This artifact can be traced back to a dynamical detuning brought about by the adiabatic XC functional.\cite{fuks+2011prb} The resonance frequency changes in time, away from the carrier frequency of the applied laser. As a result, there is an amount of excited-state population beyond which the simulation becomes inaccurate, as the GS correlation potential eventually poses a poor approximation to the dynamical one. Decreasing the laser intensity lowers this threshold, although it is reached at a later point in time (Fig. S10). Meaningful results can be expected only when staying below this threshold, \textit{i.e.} when short and weak laser pulses are considered. The pulses used in this work satisfy this condition: The effective pulse length amounts to about 5 fs, such that the field is terminated well before the problematic regime (Fig. S9).

\subsection{Transient absorption spectra}\label{tas.sec}

We next investigate transient absorption spectra, which illustrate the temporal change of the absorption characteristics of a system during and after laser irradiation. 
For ethylene and benzene, we use the same pulses as before, while for thiophene, we increase the intensity by a factor of 10 to reach a similar amount of excited-state population as in the other two molecules. 

When only the electronic part of the dynamics is considered [see Fig. \ref{tas.fig}a-c)], changes in the TAS mainly occur in the time window in which the field is active. 
In ethylene, the resonance energy is 7.6~eV, in thiophene  5.6~eV, and in benzene 6.9~eV, as marked in Fig.~\ref{spectra.fig}. Since we consider low-lying excitations, we do not observe negative features in the TAS. 
A common trait in the spectra of the three molecules is the red-shift of the pumped excitation probed during the pulse. 
This characteristic persists also after the field is switched off and is likely an artifact of the adiabatic approximation\citealp{fuks+prl2015,luo2016, prov+2015jctc} (see Sec. \ref{dyn_sim.sec}). 
Another general feature is the emergence of new peaks in the lower-energy region of the spectra during laser irradiation. 
The energies of these pre-peaks remain constant after the field is switched off, as they should.\citealp{perf+2015pra,fuks+prl2015}
On the other hand, their absorption strength is either unchanged over time, as in the case of ethylene [Fig. \ref{tas.fig}a)], or has an oscillatory time dependence, as in the other two molecules [Fig. \ref{tas.fig}b-c)]. 
These oscillations are related to the quantum beats resulting from the system being in a coherent superposition of electronic states including more than one excitation.\citealp{perf+2015pra}

When taking into account the nuclear motion [Fig. \ref{tas.fig}d-f)], the time evolution is no longer trivial after the end of the pulse. 
The absorption energies vary in time also beyond the duration of the laser illumination. 
In all cases, the pumped peak exhibits a more pronounced red-shift than the one observed when the nuclei are clamped.
This behavior can be explained in terms of the Stokes shift, namely to the red-shift due to the change of the geometry upon electronic excitation.
Unfortunately, we are unable to quantify this effect, as the corresponding spectral signatures are superimposed with the red-shift of the pumped peak discussed above for the dynamics with fixed ions. 

The behavior of the emerging pre-peaks varies depending on the system. 
In the TAS of ethylene, the distinct peak at 2~eV in Fig. \ref{tas.fig}d) exhibits an oscillatory blue-shift, which mirrors the red-shift of the pumped peak. 
In thiophene, the first pre-peak at about 1.5 eV [Fig. \ref{tas.fig}e)] is broadened in energy around 30 fs before narrowing and red-shifting. 
The second peak at about 3 eV remains instead largely unchanged, and the third one at approximately 4.5 eV exhibits a blue-shift similar to that observed in the TAS of ethylene. 
Finally, in benzene [Fig. \ref{tas.fig}f)] the first pre-peak  around 1 eV disappears completely after 30 fs, whereas the next three maxima at 2 eV, 3 eV, and 4 eV become more defined and gain oscillator strength with time in comparison to their dynamics with fixed nuclei.

\subsection{Excited-state absorption}
\label{sec:esa}
\begin{figure}
	\includegraphics[width=.45\textwidth]{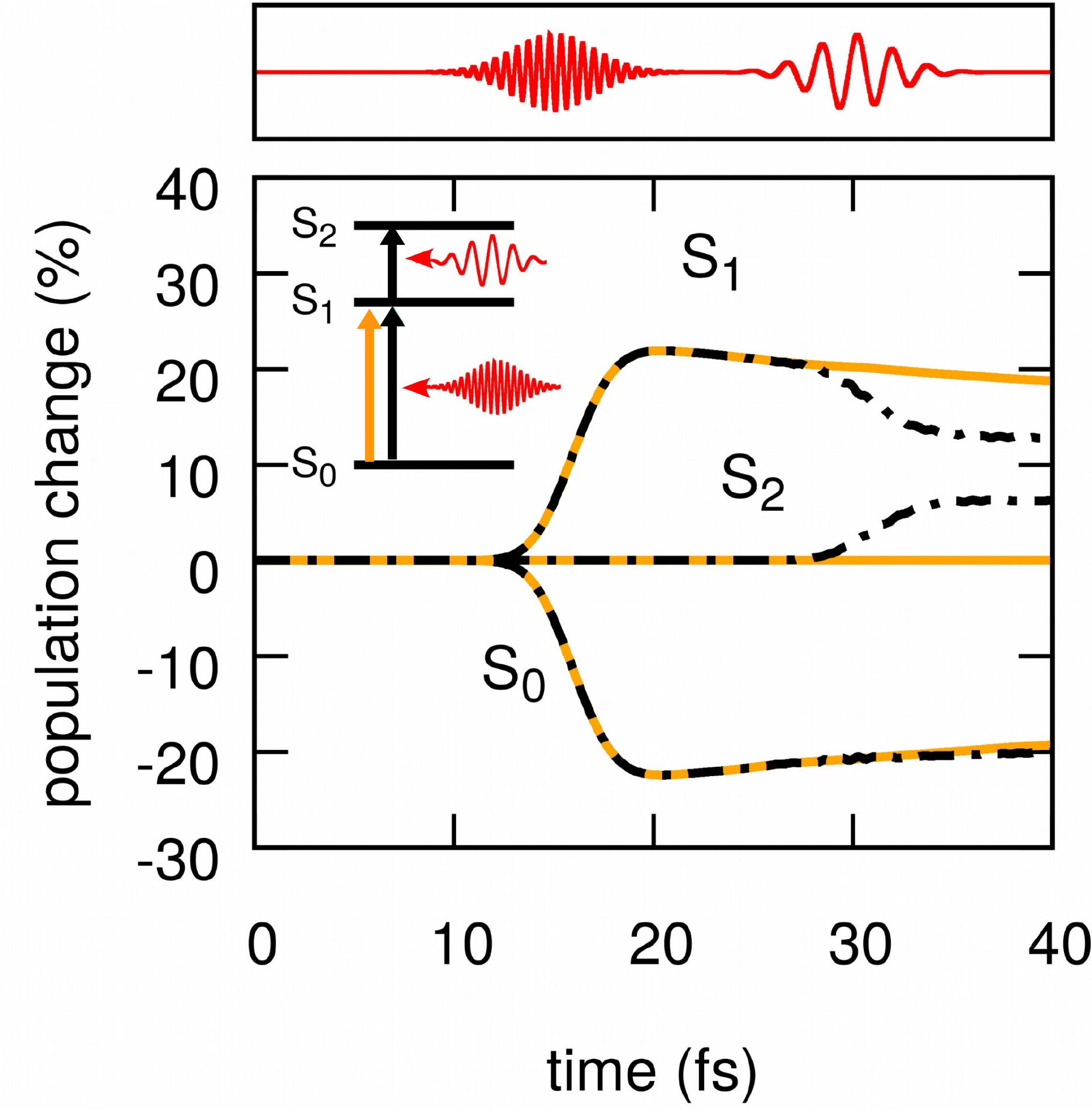}
	\caption{Variations of the populations of the GS ($S_0$) as well as of excited states ($S_1$ and $S_2$) of ethylene, computed from Eq.~\eqref{pop_single.eq} and smoothed out, upon the application of the two laser pulses sketched in the top panel. The first one is tuned to the first $\pi\rightarrow\pi^*$ transition, the second one has a frequency corresponding to the energy of the prepeak appearing in the TAS (Fig. \ref{tas.fig}). The solid curve in orange is obtained applying only the first pulse depicted in the uppermost panel, while the dotted-dashed one in black is computed when the two pulses are applied subsequently, as illustrated in the inset.}
	\label{twopulse.fig}
\end{figure}

To better understand the nature of the pre-peaks appearing in the TAS (Fig. \ref{tas.fig}), we analyze in more details the feature around 2.3 eV in the spectrum of ethylene [Fig. \ref{tas.fig}a)]. 
To do so, we interrogate the system by applying a \textit{monochromatic} probe pulse in resonance with the excitation energy of the pre-peak (2.3 eV), after the excitation with the first resonant pump at 7.6 eV. 
The total field in this case is a pulse train, depicted in the upper panel of Fig. \ref{twopulse.fig}. 
To understand the resulting electron dynamics, we monitor the populations of the individual orbitals contributing to the total number of excited electrons:
\begin{equation}
    p_m(t) = 2\sum\limits_j^{\text{occ}}\left|\langle\phi_m(0)|\phi_j(t)\rangle\right|^2.
    \label{pop_single.eq}
\end{equation}
Upon the application of the pumping laser, we find mainly the LUMO gaining population, but also a higher unoccupied state participating. 
As these orbitals contribute to one single excitation, we add them up and consider their sum as the occupation of the excited state at 7.6 eV, which we name $S_1$. We recall, however, that based on our ALDA results shown in Fig. \ref{spectra.fig}a), this is not the lowest-energy excitation of ethylene.

The increase of the population of $S_1$ is accompanied by a symmetric decrease of the population of $S_0$, which is calculated from the population of the static HOMO only, since the TD HOMO is the only active orbital. 
The following gradual decrease of the $S_1$ population is likely related to an artifact of the adiabatic appoximation and can be alleviated by choosing a weaker pump intensity (see discussion in Section S4 in the Supporting Information).
When the probe pulse is applied, the populations of the orbitals contributing to $S_1$ decrease significantly, while three higher-energy orbitals gain a corresponding amount of population.
Assuming that these three single-particle states are associated with one single ES, we add them and identify their sum as the population of state $S_2$ (details in the Supporting Information, Section S3). 
Notably, the population of $S_0$ is not affected by the probe pulse. 
We thus conclude that the pre-peak in the TAS is related to excited-state absorption from $S_1$ to $S_2$ about 2.3 eV higher in energy, as previously speculated in Ref.~\citenum{giov+2013chpch}. Overall, it is a nontrivial task to assign the states that are related to the excited-state absorption features. A general indication can be obtained from the excited-state energies from a linear-response calculation (see Supporting Information, Section S3 and Fig. S6).

\begin{figure}
\includegraphics[width=.45\textwidth]{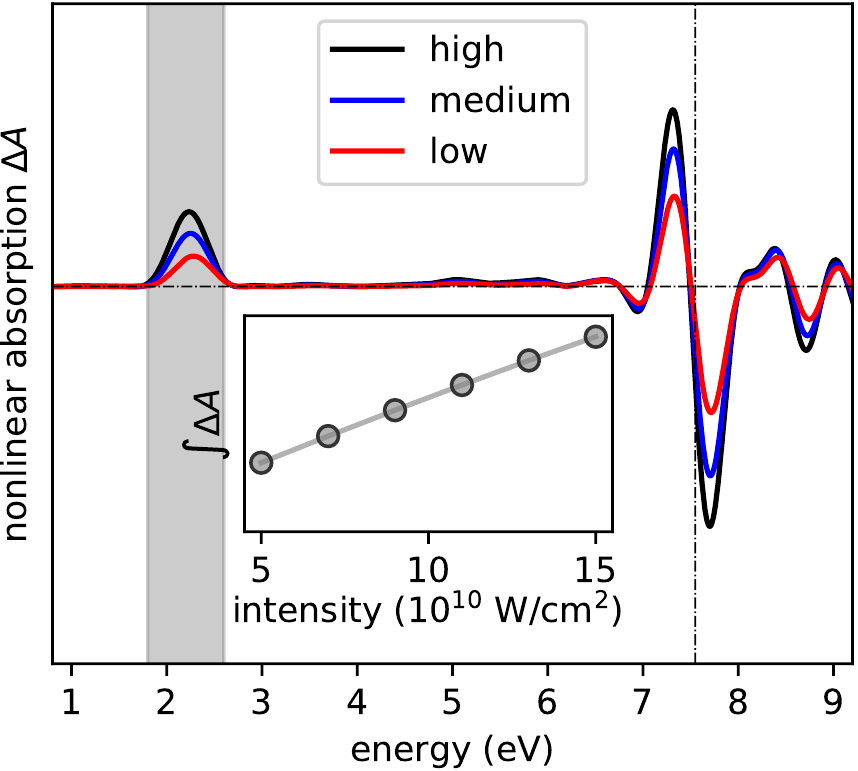}
	\caption{
	Transient absorption spectrum of ethylene right after the pulse with intensities $5\times 10^{10}$ W/cm$^2$ (low), $9\times 10^{10}$ W/cm$^2$ (medium), and $13\times 10^{10}$ W/cm$^2$ (high). The linear absorption spectrum has been subtracted in all cases. The vertical dashed line marks the carrier frequency of the laser. The inset shows the oscillator strength of the excited-state absorption maximum, which is computed by integrating the spectrum over the range indicated by the gray shaded area. 
	}
	\label{intensities.fig}
\end{figure}

To gain further insight, we examine the dependence of the excited-state absorption feature on the excited-state population. To this end, we calculate the post-pulse spectra for ethylene with different pump intensities, which, in the linear regime, are proportional to the achieved excited-state population. After subtracting the GS absorption spectrum, we see that the energy of the pre-peak is robust, \textit{i.e.} it does not undergo any significant change when increasing the laser intensity (Fig. \ref{intensities.fig}). The oscillator strength of the excited-state absorption is evaluated by integrating the area below the corresponding peak over the region highlighted in gray in Fig.~\ref{intensities.fig}. It is evident from the inset of Fig. 6 that this value is linear with respect to the laser intensity and, thus, with respect to the excited-state population.
This is the correct behavior of a non-equilibrium response function.\cite{fuks+prl2015}

\subsection{Vibronic couplings}
\label{vibr.sec}

\begin{figure}
	\includegraphics[width=.45\textwidth]{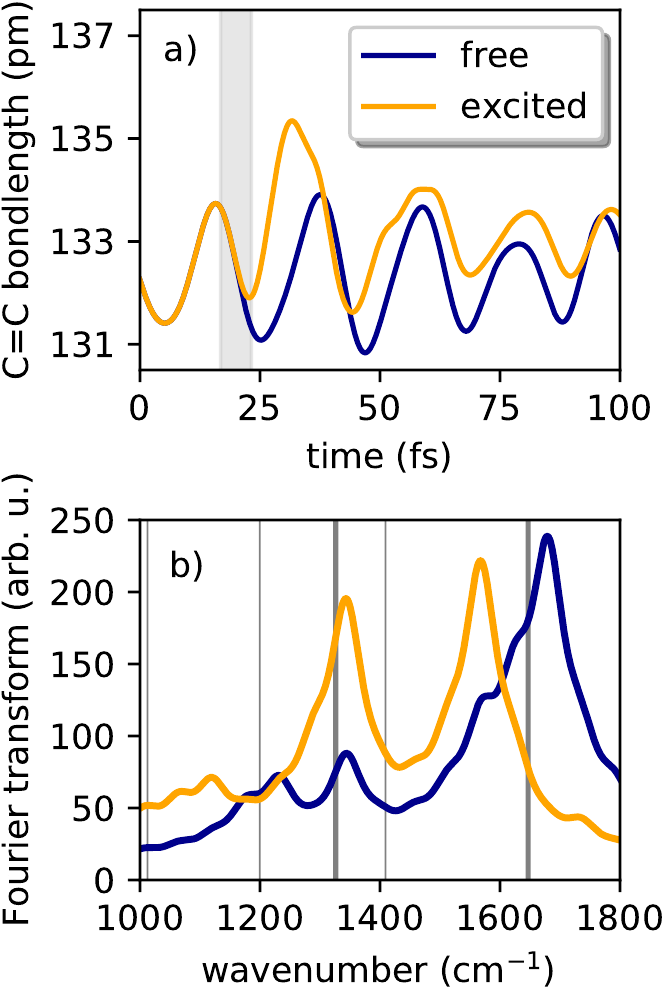}
	\caption{a) Time-dependent C=C bond length in ethylene thermally excited at 300 K obtained from a free propagation (blue) and from a propagation with a resonant pulse acting in the time window indicated by the grey area (orange). b) Time-dependent C=C bond length in frequency space obtained by Fourier-transforming the C=C bond length in panel a) propagated over 600 fs. The vertical bars denote the normal frequencies of the molecule, with the thick ones marking modes involving a change of the C=C bond length.}
	\label{bondlength.fig}
\end{figure}

After clarifying the nature of the peaks in the TAS, we now examine their dynamical behavior when the vibronic coupling is taken into account. 
In this analysis, we focus again on ethylene, since in this molecule, formed by a carbon double bond with four terminating H atoms, it is straightforward to make a connection between the temporal evolution of the absorption peaks, the vibrational properties, and the excited-state geometry. 
As we are also interested in a possible change of the vibrational modes upon electronic excitation, in this study we assign to the nuclei initial velocities, taken randomly from a Maxwell-Boltzmann distribution with a temperature of 300 K. 
We then propagate the system freely (\textit{i.e.}, without laser perturbation) for about 50 fs, to enable a uniform distribution of the kinetic energy among the nuclear degrees of freedom. 
Finally, we extract coordinates and velocities and use them as a starting point for two simulations: The first one is another free propagation, corresponding to the nuclear motion on the GS potential energy surface (PES); in the second calculation, the molecule is excited with a laser pulse targeting the HOMO-LUMO transition at 7.6 eV. 
In the latter, the dynamics takes place on both the GS PES and the ES PES. 
In both cases, we monitor the time evolution of the C=C bond length, plotted in Fig. \ref{bondlength.fig}a). 

From inspection of Fig. \ref{bondlength.fig}a), we immediately notice that the electronically excited molecule displays a general increase of the  C=C bond length. 
This behavior can be understood in more detail from the induced density [Eq. \eqref{delta_rho_model.eq}] and from the difference density $\Delta\rho^{(e)}(\textbf{r})$ of the pumped excitation computed in the linear-response framework [Eq. \eqref{static_ind_dens.eq}]. 
According to Eq. \eqref{delta_rho_model.eq}, the induced density $\delta\rho(\textbf{r},t)$ is the sum of two terms: The first one is related to $\Delta\rho^{(e)}(\textbf{r})$ and changes relatively slowly ($\sim$10 fs for the considered laser intensity), and only while the field is active. 
The density redistribution related to $\Delta\rho^{(e)}(\textbf{r})$ causes static forces on the positively charged nuclei, driving them away from their equilibrium positions. 
For the specific case of ethylene, the electron depletion along the C=C bond creates a positive charge in the center of the molecule that repels the positively charged nuclei, while the electron accumulation on the sides causes attractive forces [see inset of Fig. \ref{spectra.fig}a)]. 
As a result, the carbon atoms are driven apart and the C=C bond is elongated. 
The second contribution to $\delta\rho(\textbf{r},t)$ is related to the transition density $\rho_t^{(e)}(\textbf{r})$ [Eq.~\eqref{transition_density.eq}] of the electronic excitation and corresponds to the coherent electronic motion. 
The resulting forces on the nuclei, however, have a strong oscillatory time dependence ($<$ 1 fs, determined by the electronic excitation frequency), and average to zero. 
As such, they cannot trigger the nuclear motion effectively. 
Hence, the difference density $\Delta\rho^{(e)}(\textbf{r})$ provides us with more information about the vibronic coupling than the transition density $\rho_t^{(e)}(\textbf{r})$.

In addition to the laser-induced elongation of the C=C bond length discussed above, in Fig. \ref{bondlength.fig}a) we also notice another striking feature triggered by the pulse.
While in the free propagation the C=C bond length exhibits an almost harmonic oscillation in time, right after the laser irradiation, between 25 fs and 50 fs, we find a maximum formed by a double peak. 
This feature suggests that upon laser illumination the motion of the C=C bond becomes a superposition of two oscillations with different frequencies. 
To investigate this point further, we perform a Fourier transform of both curves shown in Fig. \ref{bondlength.fig}a) after extending the propagation up to 600 fs, in order to obtain a spectrum with sufficient frequency resolution [see Fig.~\ref{bondlength.fig}b)]. 
The free propagation yields a pronounced peak close to the frequency of the C=C stretching mode (1650~cm$^{-1}$), in agreement with the harmonic behavior observed in time domain [Fig.~\ref{bondlength.fig}a)]. 
Another local maximum is recognizable in the vicinity of the C=C scissoring mode around 1350~cm$^{-1}$.
When the molecule is electronically excited, this peak undergoes a red-shift of about 150~cm$^{-1}$, which we attribute to the anharmonicity of the oscillation. 
The most striking feature in the orange curve in Fig.~\ref{bondlength.fig}b) is, however, the dramatic increase in strength of the peak at about 1350~cm$^{-1}$ as a consequence of vibronic coupling. The frequency of this peak remains unchanged regardless of the laser perturbation, suggesting low anharmonicity in this motion, likely due to the small amplitude of the oscillation prior to the electronic excitation. 
The presence of two maxima with similar oscillator strength in the spectrum of the laser-excited molecule [Fig.~\ref{bondlength.fig}b] clarifies the double peak in Fig.~\ref{bondlength.fig}a) as a superposition of two oscillations in the time domain. 

The displayed behavior of the vibrational motion in ethylene depends on the specific initial conditions, and the same electric field will cause a different response when the molecule is in a different nuclear configuration. 
For a more general analysis, a statistical approach is required, performing the same calculations for a number of different starting conditions. 
The presented results correspond to a particular scenario but show significant effects that a laser excitation can exert with respect to the vibrational properties. 
Through the vibronic coupling, the electronic excitation energy is partially transferred to the nuclei, increasing the energy in the normal modes. 
Meanwhile, the modes can exhibit a frequency shift, which is likely related to the anharmonicity of the potential energy associated with the normal modes.

\begin{figure*}
    \centering
\includegraphics[width = \textwidth]{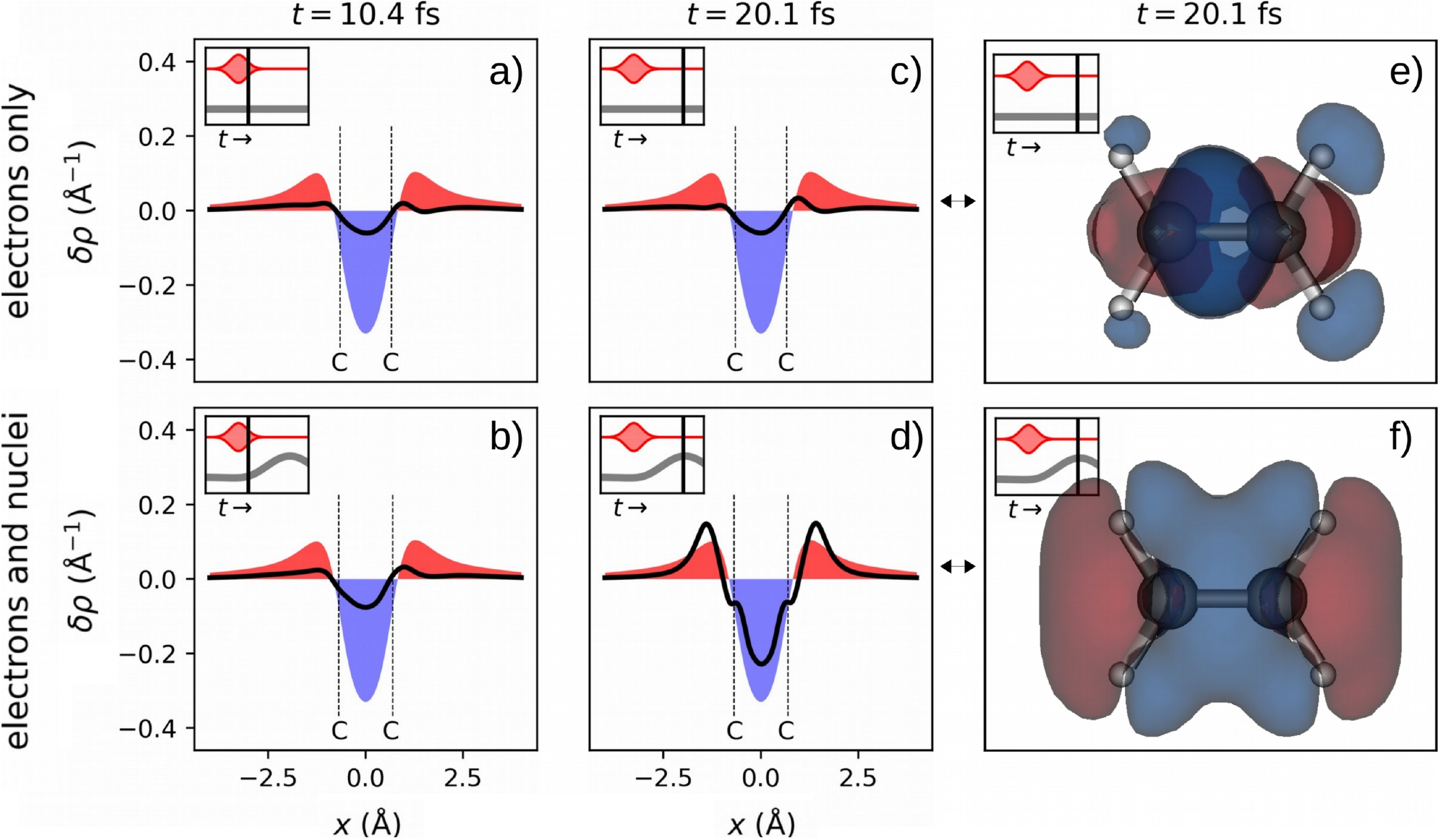}
    \caption{Time-dependent induced density $\delta\rho(x,t)$ in ethylene (black lines) at $t$=10.4 fs, i.e. immediately after laser application [panels a) and b)], and $t=20.1$~fs, corresponding to the maximum of the C=C bond length [panels c) and d)]. The insets show the laser envelope and the C=C bond length versus time, with the vertical line indicating the instant of the snapshot. The filled colored areas in the background of panels a)-d) illustrate the difference density $\Delta\rho^{(e)}(x)$ of the excitation, computed in linear response. Panels e) and f) show the isosurfaces of $\delta\rho(\textbf{r}, t=20.1\,\text{fs})$ with isovalues of $\pm 3\times 10^{-3}\,\text{\AA}^{-3}$. In the upper panels a), c), and e), the nuclei are kept fixed, while in b), d), and f), they are free to move.}
    \label{ind_dens.fig}
\end{figure*}

To conclude this analysis of the electron-nuclear dynamics of ethylene, we inspect the feedback of the nuclear motion to the electronic excitation. 
To do so, in Fig. \ref{ind_dens.fig} we plot the time-dependent induced density computed for fixed and moving ions. 
We notice that, immediately after the excitation, the inclusion of the nuclear motion generates only a slight difference in the induced density compared to the purely electronic contribution [Fig. \ref{ind_dens.fig}a) and b)]: The induced electronic density, integrated over the plane perpendicular to the bond axis [see coordinate system in Fig. \ref{spectra.fig}a)],
\begin{equation}
\delta\rho(x,t) = \int\text dy\text dz\,\left[\rho(\textbf{r},t)-\rho_g(\textbf{r})\right],
\end{equation}
is very similar in the case of fixed and moving nuclei. 
Towards the end of the pulse, the electrostatic forces on the nuclei due to the laser-induced electronic redistribution become noticeable. 
When the nuclei are allowed to move, the carbon atoms are driven apart and the bond length increases [inset of Fig. \ref{ind_dens.fig}b)]. 
After approximately 10 fs, the bond length reaches a maximum and $\delta\rho(x,t)$ deviates significantly from the case in which the nuclei are fixed [Fig. \ref{ind_dens.fig}c) and d)]: The electron depletion between the carbon atoms and the accumulation on the outside are enhanced considerably when the nuclei are allowed to move [Fig. \ref{ind_dens.fig}d)]. 
When the nuclei are clamped, $\delta\rho(\textbf{r},t)$ looks similar at t $\simeq$ 10 fs and t $\simeq$ 20 fs [Fig. \ref{ind_dens.fig}a) and c)], since the electronic system propagates freely after the excitation. 
We compare $\delta\rho(x,t)$ to the plane-integrated difference density of the excitation,
\begin{equation}
\Delta\rho^{(e)}(x) = \int\text dy\text dz\,\Delta\rho^{(e)}(\textbf{r})= \int\text dy\text dz\,\left[\rho_e(\textbf{r})-\rho_g(\textbf{r})\right],
\end{equation}
where $\Delta\rho^{(e)}(\textbf{r})$ is calculated in linear response from Eq. \eqref{static_ind_dens.eq}. 
This is the value that $\delta\rho(\textbf{r},t)$ would assume if the laser evoked 100\% ES population. 
As this scenario does no longer represent a superposition state, the density is stationary.  
When the nuclear motion is included, $\delta\rho(x,t)$ shows a clear bias toward $\Delta\rho^{(e)}(x)$ at 20 fs [Fig. \ref{ind_dens.fig}d)]. 
This is evident when comparing the corresponding three-dimensional representations of the induced density [Fig. \ref{ind_dens.fig}e-f)] to the one given in the inset of Fig. \ref{spectra.fig}a): The nuclear motion strongly enhances the charge redistribution of the excitation. 
It should be emphasized that at $t=20.1$ fs the C=C bond length exhibits a maximum. 
In the subsequent nuclear oscillations, the electron density continuously adapts to the vibrational phase, and in the minima of the bond length the enhancement is substantially weaker. 
However, since the mean bond length expands [Fig. \ref{bondlength.fig}a)], there is a net increase of electron localization that remains after averaging over a vibrational cycle.

\begin{figure}
\includegraphics[width=.5\textwidth]{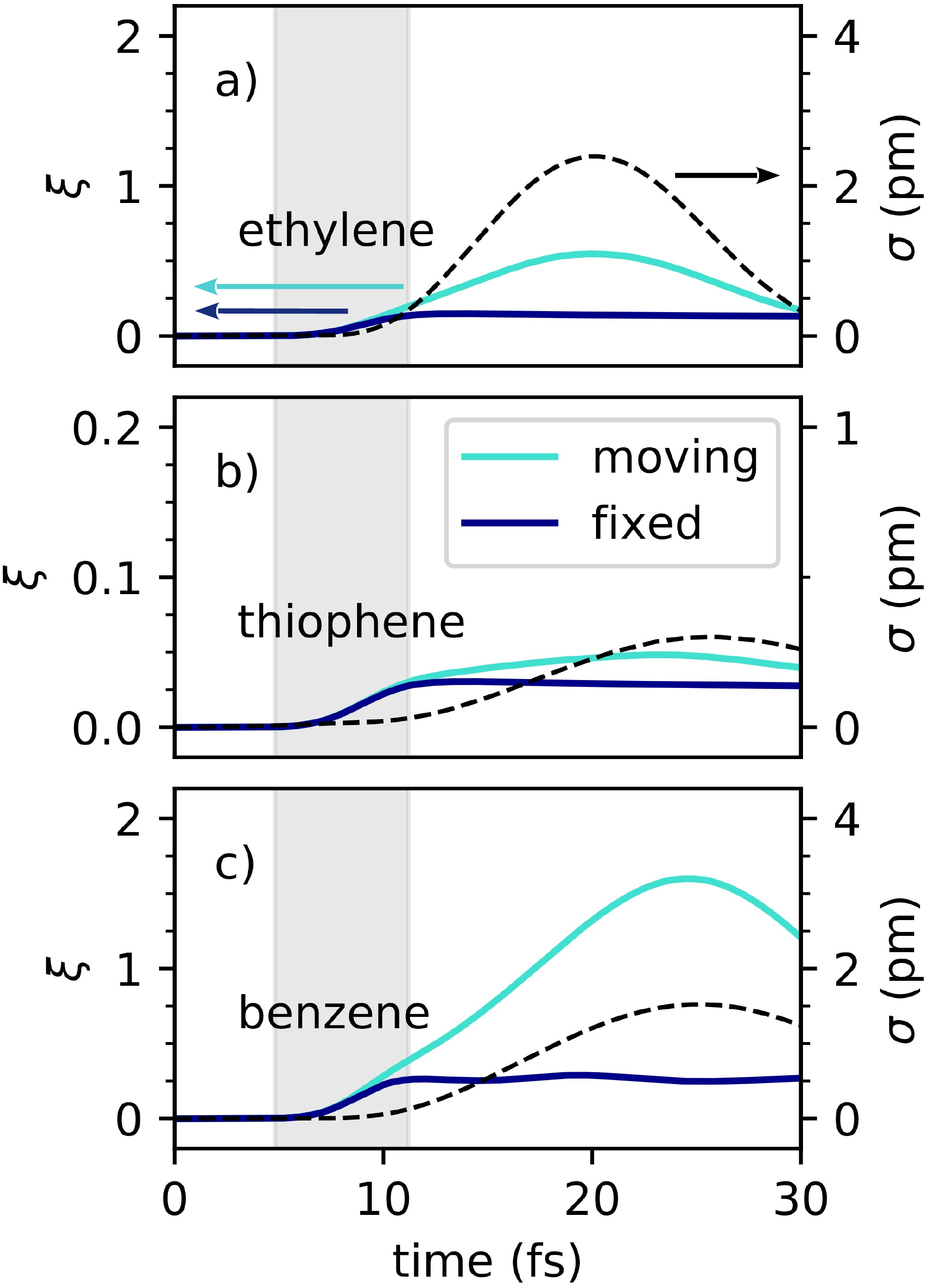}
	\caption{$\xi(t)$, computed from Eq. \eqref{xi.eq}, for ethylene (a), thiophene (b), and benzene (c) with moving (cyan solid line) and fixed nuclei (blue solid line). The black dashed lines represent $\sigma$, namely the deviation of the geometry from the equilibrium one [Eq. \eqref{sigma.eq}]. The grey shaded area indicates the active time window of the laser.}
	\label{xi.fig}
\end{figure}

For a more quantitative and transferable analysis of the electron localization caused by the moving nuclei, we define a function measuring the similarity between the induced density $\delta\rho(\textbf{r},t)$ and the difference density $\Delta\rho^{(e)}(\textbf{r})=\rho_e(\textbf{r})-\rho_g(\textbf{r})$ of the considered excitation:
\begin{equation}\label{xi.eq}
    \xi(t) = \frac{\int\text dt'\, w(t-t')\int\text d^3r\,\delta\rho(\textbf{r},t')\Delta\rho^{(e)}(\textbf{r)}}{\int\text d^3r\,\Delta\rho^{(e)}(\textbf{r)}^2},
\end{equation}
where $w(t)$ is a normalized window function with an extension such that the strong oscillations related to the coherent electronic motion are removed, but the trend of the curves is maintained. 
$\xi(t)$ is a dimensionless quantity corresponding to the excited-state population when the nuclei are kept fixed (further details in the Supporting Information, Section S6). 
More generally, $\xi(t)$ measures the similarity between the TD density and the density of the vertically excited state. 
If the nuclei are kept fixed, only an electronic excitation contributes to $\xi(t)$; if they are released, the nuclear relaxation can cause an corresponding redistribution of charge. 
$\xi(t)=0$ indicates that the density displacements are uncorrelated with the excitation.
Conversely, if $\xi(t)=1$, the electron density of the system is close to that of the molecule in a vertically excited state with 100\% ES population.
Comparing $\xi(t)$ obtained in the simulations with clamped and moving nuclei, we can determine to which extent the nuclear motion enhances the excitation-induced charge redistribution. 
We compare this result to an indicator for the deviation of the nuclear geometry from the equilibrium one:
\begin{equation}\label{sigma.eq}
    \sigma(t) = \left(\frac{\sum_J Z_J^2\left|\textbf{R}_J(t)-\textbf{R}_J(0)\right|^2}{\sum_J Z_J^2}\right)^{\frac 12}.
\end{equation}
The individual atomic displacements $|\textbf{R}_J(t)-\textbf{R}_J(0)|$ are weighted by the valence charge $Z_J$ of the corresponding pseudopotential, which is proportional to the electrostatic force exerted by the positive pseudo-nuclei and thus to the interaction strength with the electronic system. 
The quantity in Eq.~\eqref{sigma.eq} predicts quite well the positions of the maxima of $\xi(t)$ (see Fig. \ref{xi.fig}). 
In the case of ethylene [Fig. \ref{xi.fig}a)], $\sigma(t)$ essentially describes the change of the C=C bond length, as its HOMO-LUMO excitation is mainly coupled to vibrational modes involving a C=C stretch (see discussion above). 
Both $\xi(t)$ and $\sigma(t)$ have a maximum at around 20 fs, with $\xi(t)$ exhibiting a peak at about 0.6 when the nuclei are in motion, in contrast to the constant value of 0.2 assumed after the laser illumination when the nuclei are clamped (see Fig. \ref{ind_dens.fig}). 
Also in the case of thiophene, the nuclear motion enhances the value of $\xi(t)$ [Fig. \ref{xi.fig}b)]. 
The difference between fixed and moving nuclei is however less pronounced than in ethylene, which we attribute to the larger mass of S in comparison to H and C. 
The most pronounced discrepancy between fixed and moving nuclei is observed in benzene, for which $\xi(t)$ assumes a maximum value of about 1.5 when nuclei are moving, compared to 0.4 with fixed nuclei [Fig. \ref{xi.fig}c)]. 
The fact that $\xi(t)$ exceeds 1 in the former case is likely related to the fact that the pumped excitation is degenerate (see also Supprting Information, Table S3).

To rationalize the increase of $\xi(t)$ upon the inclusion of the nuclear motion, we first note that the vibrational energies are insufficient to cause interband transitions. 
Thus, the ES population cannot be increased by the nuclear motion. 
We instead attribute the enhancement effect to the change of the GS electron density associated with a change of the nuclear configuration. 
The electronic excitation leads to a partial population of the excited-state PES in a point which is not a minimum, such that forces begin to drive the nuclear configuration away from the GS equilibrium one ($R$). 
The new configuration ($R+\Delta R$) has a different GS density, which is similar to that of the vertically excited state, \textit{i.e.}, $\rho_g(R+\Delta R)\sim \rho_e(R)$. 
This relaxation-related charge redistribution acts on top of the one stemming directly from the electronic excitation. 
As shown in Fig. \ref{xi.fig}, the ratio between these two contributions depends on the system: In thiophene, the electronic excitation causes the most pronounced change in the density; in ethylene and particularly in benzene, this role is played by the nuclear motion.

\section{Summary and Conclusions}
\label{sec:conclu}
We have investigated the laser-induced dynamics of three small organic molecules (ethylene, benzene, and thiophene) by means of RT-TDDFT in the adiabatic local density approximation. 
The scope of our study was to introduce simple models and indicators to interpret the quantities computed with this method, and to better understand the physics of ultrafast dynamical processes.
By exciting each molecule with a laser pulse in resonance with a low-lying electronic excitation, we have monitored the time evolution of the induced dipole moment and of the electronic population. 
The comparison with the results obtained from an exactly-solvable two-level model has provided a diagnostic tool to analyze the computed quantities and to interpret the corresponding results in a physically transparent manner. 
The analysis of the transient absorption spectra, computed with and without including the nuclear motion, has revealed the complex dynamics of laser-excited systems, even of such small size as those examined in this work. 
The appearance of excited-state absorption, induced by the laser pulse and probed at subsequent time delays, has been demonstrated by an \textit{ad hoc} simulation triggered by a train of monochromatic pulses. 
Finally, the analysis of the coupled electron-nuclear motion has allowed us to disclose the effects of the external field on the vibrational degrees of freedom, and to reveal the crucial role of vibronic coupling in stabilizing the excitation already in the first few fs after the laser illumination.
These effects have been shown by a visual representation of the induced charge density distribution as well as through general indicators applicable also to more complex molecular systems.

This work and the analysis reported therein offer useful tools to understand the results of RT-TDDFT in describing ultrafast dynamical processes within the given approximations. 
We expect that this contribution will further consolidate the role of RT-TDDFT coupled to the Ehrenfest molecular dynamics scheme as state-of-the-art methodology capable to address the fascinating and yet complex phenomena of light-matter interaction in the femtosecond regime.

\section*{Supporting Information}
The Supporting Information includes further details about the linear-response calculations of the relevant excitations, a methodological section related to the excited-state absorption calculations of Sec.~\ref{dyn_sim.sec}, the analysis of the excited-state spectra of thiophene, additional computational details about the two-level model, a section discussing the effect of adding a self-interaction correction on nonequilibrium spectra and population dynamics, and different approaches of calculating the the excited-state population from RT-ALDA. Ref.~\onlinecite{pbe} is additionally cited therein.


\section*{Acknowledgement}
The authors acknowledge fruitful discussions with Antonietta de Sio and Alberto Guandalini. This work was funded by the Deutsche Forschungsgemeinschaft (DFG, German Research Foundation) -- Project number 182087777 -- SFB 951 and Project number 86798544 -- HE 5866/2-1.
Computational resources are provided by the North-German Supercomputing Alliance (HLRN), project bep00060.


\section*{Data availability}
The data that support the findings of this study are available from the authors upon reasonable request.


\begin{thebibliography}{87}%
\makeatletter
\providecommand \@ifxundefined [1]{%
 \@ifx{#1\undefined}
}%
\providecommand \@ifnum [1]{%
 \ifnum #1\expandafter \@firstoftwo
 \else \expandafter \@secondoftwo
 \fi
}%
\providecommand \@ifx [1]{%
 \ifx #1\expandafter \@firstoftwo
 \else \expandafter \@secondoftwo
 \fi
}%
\providecommand \natexlab [1]{#1}%
\providecommand \enquote  [1]{``#1''}%
\providecommand \bibnamefont  [1]{#1}%
\providecommand \bibfnamefont [1]{#1}%
\providecommand \citenamefont [1]{#1}%
\providecommand \href@noop [0]{\@secondoftwo}%
\providecommand \href [0]{\begingroup \@sanitize@url \@href}%
\providecommand \@href[1]{\@@startlink{#1}\@@href}%
\providecommand \@@href[1]{\endgroup#1\@@endlink}%
\providecommand \@sanitize@url [0]{\catcode `\\12\catcode `\$12\catcode
  `\&12\catcode `\#12\catcode `\^12\catcode `\_12\catcode `\%12\relax}%
\providecommand \@@startlink[1]{}%
\providecommand \@@endlink[0]{}%
\providecommand \url  [0]{\begingroup\@sanitize@url \@url }%
\providecommand \@url [1]{\endgroup\@href {#1}{\urlprefix }}%
\providecommand \urlprefix  [0]{URL }%
\providecommand \Eprint [0]{\href }%
\providecommand \doibase [0]{http://dx.doi.org/}%
\providecommand \selectlanguage [0]{\@gobble}%
\providecommand \bibinfo  [0]{\@secondoftwo}%
\providecommand \bibfield  [0]{\@secondoftwo}%
\providecommand \translation [1]{[#1]}%
\providecommand \BibitemOpen [0]{}%
\providecommand \bibitemStop [0]{}%
\providecommand \bibitemNoStop [0]{.\EOS\space}%
\providecommand \EOS [0]{\spacefactor3000\relax}%
\providecommand \BibitemShut  [1]{\csname bibitem#1\endcsname}%
\let\auto@bib@innerbib\@empty
\bibitem [{\citenamefont {Andrade}\ \emph {et~al.}(2012)\citenamefont
  {Andrade}, \citenamefont {Alberdi-Rodriguez}, \citenamefont {Strubbe},
  \citenamefont {Oliveira}, \citenamefont {Nogueira}, \citenamefont {Castro},
  \citenamefont {Muguerza}, \citenamefont {Arruabarrena}, \citenamefont
  {Louie}, \citenamefont {Aspuru-Guzik}, \citenamefont {Rubio},\ and\
  \citenamefont {Marques}}]{andr+jpcm2012}%
  \BibitemOpen
  \bibfield  {author} {\bibinfo {author} {\bibfnamefont {X.}~\bibnamefont
  {Andrade}}, \bibinfo {author} {\bibfnamefont {J.}~\bibnamefont
  {Alberdi-Rodriguez}}, \bibinfo {author} {\bibfnamefont {D.~A.}\ \bibnamefont
  {Strubbe}}, \bibinfo {author} {\bibfnamefont {M.~J.~T.}\ \bibnamefont
  {Oliveira}}, \bibinfo {author} {\bibfnamefont {F.}~\bibnamefont {Nogueira}},
  \bibinfo {author} {\bibfnamefont {A.}~\bibnamefont {Castro}}, \bibinfo
  {author} {\bibfnamefont {J.}~\bibnamefont {Muguerza}}, \bibinfo {author}
  {\bibfnamefont {A.}~\bibnamefont {Arruabarrena}}, \bibinfo {author}
  {\bibfnamefont {S.~G.}\ \bibnamefont {Louie}}, \bibinfo {author}
  {\bibfnamefont {A.}~\bibnamefont {Aspuru-Guzik}}, \bibinfo {author}
  {\bibfnamefont {A.}~\bibnamefont {Rubio}}, \ and\ \bibinfo {author}
  {\bibfnamefont {M.~A.~L.}\ \bibnamefont {Marques}},\ }\bibfield  {title}
  {\enquote {\bibinfo {title} {Time-dependent density-functional theory in
  massively parallel computer architectures: the octopus project},}\
  }\href@noop {} {\bibfield  {journal} {\bibinfo  {journal}
  {J.~Phys.~Condens.~Matter.~}\ }\textbf {\bibinfo {volume} {24}},\ \bibinfo
  {pages} {233202} (\bibinfo {year} {2012})}\BibitemShut {NoStop}%
\bibitem [{\citenamefont {Jornet-Somoza}\ \emph {et~al.}(2015)\citenamefont
  {Jornet-Somoza}, \citenamefont {Alberdi-Rodriguez}, \citenamefont {Milne},
  \citenamefont {Andrade}, \citenamefont {Marques}, \citenamefont {Nogueira},
  \citenamefont {Oliveira}, \citenamefont {Stewart},\ and\ \citenamefont
  {Rubio}}]{10000atoms}%
  \BibitemOpen
  \bibfield  {author} {\bibinfo {author} {\bibfnamefont {J.}~\bibnamefont
  {Jornet-Somoza}}, \bibinfo {author} {\bibfnamefont {J.}~\bibnamefont
  {Alberdi-Rodriguez}}, \bibinfo {author} {\bibfnamefont {B.~F.}\ \bibnamefont
  {Milne}}, \bibinfo {author} {\bibfnamefont {X.}~\bibnamefont {Andrade}},
  \bibinfo {author} {\bibfnamefont {M.~A.~L.}\ \bibnamefont {Marques}},
  \bibinfo {author} {\bibfnamefont {F.}~\bibnamefont {Nogueira}}, \bibinfo
  {author} {\bibfnamefont {M.~J.~T.}\ \bibnamefont {Oliveira}}, \bibinfo
  {author} {\bibfnamefont {J.~J.~P.}\ \bibnamefont {Stewart}}, \ and\ \bibinfo
  {author} {\bibfnamefont {A.}~\bibnamefont {Rubio}},\ }\bibfield  {title}
  {\enquote {\bibinfo {title} {Insights into colour-tuning of chlorophyll
  optical response in green plants},}\ }\href {\doibase 10.1039/C5CP03392F}
  {\bibfield  {journal} {\bibinfo  {journal} {Phys. Chem. Chem. Phys.}\
  }\textbf {\bibinfo {volume} {17}},\ \bibinfo {pages} {26599--26606} (\bibinfo
  {year} {2015})}\BibitemShut {NoStop}%
\bibitem [{\citenamefont {Yabana}\ and\ \citenamefont
  {Bertsch}(1996)}]{yaba+prb1996}%
  \BibitemOpen
  \bibfield  {author} {\bibinfo {author} {\bibfnamefont {K.}~\bibnamefont
  {Yabana}}\ and\ \bibinfo {author} {\bibfnamefont {G.~F.}\ \bibnamefont
  {Bertsch}},\ }\bibfield  {title} {\enquote {\bibinfo {title} {Time-dependent
  local-density approximation in real time},}\ }\href@noop {} {\bibfield
  {journal} {\bibinfo  {journal} {Phys. Rev. B}\ }\textbf {\bibinfo {volume}
  {54}},\ \bibinfo {pages} {4484--4487} (\bibinfo {year} {1996})}\BibitemShut
  {NoStop}%
\bibitem [{\citenamefont {Cocchi}\ \emph {et~al.}(2014)\citenamefont {Cocchi},
  \citenamefont {Prezzi}, \citenamefont {Ruini}, \citenamefont {Molinari},\
  and\ \citenamefont {Rozzi}}]{cocc+prl2014}%
  \BibitemOpen
  \bibfield  {author} {\bibinfo {author} {\bibfnamefont {C.}~\bibnamefont
  {Cocchi}}, \bibinfo {author} {\bibfnamefont {D.}~\bibnamefont {Prezzi}},
  \bibinfo {author} {\bibfnamefont {A.}~\bibnamefont {Ruini}}, \bibinfo
  {author} {\bibfnamefont {E.}~\bibnamefont {Molinari}}, \ and\ \bibinfo
  {author} {\bibfnamefont {C.~A.}\ \bibnamefont {Rozzi}},\ }\bibfield  {title}
  {\enquote {\bibinfo {title} {Ab initio simulation of optical limiting: The
  case of metal-free phthalocyanine},}\ }\href@noop {} {\bibfield  {journal}
  {\bibinfo  {journal} {Phys.~Rev.~Lett.~}\ }\textbf {\bibinfo {volume}
  {112}},\ \bibinfo {pages} {198303} (\bibinfo {year} {2014})}\BibitemShut
  {NoStop}%
\bibitem [{\citenamefont {De~Giovannini}\ \emph {et~al.}(2013)\citenamefont
  {De~Giovannini}, \citenamefont {Brunetto}, \citenamefont {Castro},
  \citenamefont {Walkenhorst},\ and\ \citenamefont {Rubio}}]{giov+2013chpch}%
  \BibitemOpen
  \bibfield  {author} {\bibinfo {author} {\bibfnamefont {U.}~\bibnamefont
  {De~Giovannini}}, \bibinfo {author} {\bibfnamefont {G.}~\bibnamefont
  {Brunetto}}, \bibinfo {author} {\bibfnamefont {A.}~\bibnamefont {Castro}},
  \bibinfo {author} {\bibfnamefont {J.}~\bibnamefont {Walkenhorst}}, \ and\
  \bibinfo {author} {\bibfnamefont {A.}~\bibnamefont {Rubio}},\ }\bibfield
  {title} {\enquote {\bibinfo {title} {Simulating pump-probe photoelectron
  and absorption spectroscopy on the attosecond timescale with time-dependent
  density functional theory},}\ }\href@noop {} {\bibfield  {journal} {\bibinfo
  {journal} {Chem.~Phys.~Chem.~}\ }\textbf {\bibinfo {volume} {14}},\ \bibinfo
  {pages} {1363--1376} (\bibinfo {year} {2013})}\BibitemShut {NoStop}%
\bibitem [{\citenamefont {Lopata}\ and\ \citenamefont
  {Govind}(2011)}]{lopa+jctc2011}%
  \BibitemOpen
  \bibfield  {author} {\bibinfo {author} {\bibfnamefont {K.}~\bibnamefont
  {Lopata}}\ and\ \bibinfo {author} {\bibfnamefont {N.}~\bibnamefont
  {Govind}},\ }\bibfield  {title} {\enquote {\bibinfo {title} {Modeling fast
  electron dynamics with real-time time-dependent density functional theory:
  Application to small molecules and chromophores},}\ }\href {\doibase
  10.1021/ct200137z} {\bibfield  {journal} {\bibinfo  {journal}
  {J.~Chem.~Theory~Comput.~}\ }\textbf {\bibinfo {volume} {7}},\ \bibinfo
  {pages} {1344--1355} (\bibinfo {year} {2011})}\BibitemShut {NoStop}%
\bibitem [{\citenamefont {Yamada}\ and\ \citenamefont
  {Yabana}(2019)}]{yama+2019prb}%
  \BibitemOpen
  \bibfield  {author} {\bibinfo {author} {\bibfnamefont {A.}~\bibnamefont
  {Yamada}}\ and\ \bibinfo {author} {\bibfnamefont {K.}~\bibnamefont
  {Yabana}},\ }\bibfield  {title} {\enquote {\bibinfo {title} {Multiscale
  time-dependent density functional theory for a unified description of
  ultrafast dynamics: Pulsed light, electron, and lattice motions in
  crystalline solids},}\ }\href@noop {} {\bibfield  {journal} {\bibinfo
  {journal} {Phys. Rev. B}\ }\textbf {\bibinfo {volume} {99}},\ \bibinfo
  {pages} {245103} (\bibinfo {year} {2019})}\BibitemShut {NoStop}%
\bibitem [{\citenamefont {Sato}\ \emph {et~al.}(2014)\citenamefont {Sato},
  \citenamefont {Yabana}, \citenamefont {Shinohara}, \citenamefont {Otobe},\
  and\ \citenamefont {Bertsch}}]{sato+2014prb}%
  \BibitemOpen
  \bibfield  {author} {\bibinfo {author} {\bibfnamefont {S.~A.}\ \bibnamefont
  {Sato}}, \bibinfo {author} {\bibfnamefont {K.}~\bibnamefont {Yabana}},
  \bibinfo {author} {\bibfnamefont {Y.}~\bibnamefont {Shinohara}}, \bibinfo
  {author} {\bibfnamefont {T.}~\bibnamefont {Otobe}}, \ and\ \bibinfo {author}
  {\bibfnamefont {G.~F.}\ \bibnamefont {Bertsch}},\ }\bibfield  {title}
  {\enquote {\bibinfo {title} {Numerical pump-probe experiments of
  laser-excited silicon in nonequilibrium phase},}\ }\href@noop {} {\bibfield
  {journal} {\bibinfo  {journal} {Phys. Rev. B}\ }\textbf {\bibinfo {volume}
  {89}},\ \bibinfo {pages} {064304} (\bibinfo {year} {2014})}\BibitemShut
  {NoStop}%
\bibitem [{\citenamefont {Jiao}\ \emph {et~al.}(2013)\citenamefont {Jiao},
  \citenamefont {Wang}, \citenamefont {Hong}, \citenamefont {Su}, \citenamefont
  {Chen},\ and\ \citenamefont {Zhang}}]{jiao+2013pla}%
  \BibitemOpen
  \bibfield  {author} {\bibinfo {author} {\bibfnamefont {Y.}~\bibnamefont
  {Jiao}}, \bibinfo {author} {\bibfnamefont {F.}~\bibnamefont {Wang}}, \bibinfo
  {author} {\bibfnamefont {X.}~\bibnamefont {Hong}}, \bibinfo {author}
  {\bibfnamefont {W.}~\bibnamefont {Su}}, \bibinfo {author} {\bibfnamefont
  {Q.}~\bibnamefont {Chen}}, \ and\ \bibinfo {author} {\bibfnamefont
  {F.}~\bibnamefont {Zhang}},\ }\bibfield  {title} {\enquote {\bibinfo {title}
  {Electron dynamics in cab6 induced by one- and two-color femtosecond
  laser},}\ }\href@noop {} {\bibfield  {journal} {\bibinfo  {journal}
  {Phys.~Lett.~A.~}\ }\textbf {\bibinfo {volume} {377}},\ \bibinfo {pages} {823
  -- 827} (\bibinfo {year} {2013})}\BibitemShut {NoStop}%
\bibitem [{\citenamefont {Otobe}\ \emph {et~al.}(2008)\citenamefont {Otobe},
  \citenamefont {Yamagiwa}, \citenamefont {Iwata}, \citenamefont {Yabana},
  \citenamefont {Nakatsukasa},\ and\ \citenamefont {Bertsch}}]{otob+2008prb}%
  \BibitemOpen
  \bibfield  {author} {\bibinfo {author} {\bibfnamefont {T.}~\bibnamefont
  {Otobe}}, \bibinfo {author} {\bibfnamefont {M.}~\bibnamefont {Yamagiwa}},
  \bibinfo {author} {\bibfnamefont {J.-I.}\ \bibnamefont {Iwata}}, \bibinfo
  {author} {\bibfnamefont {K.}~\bibnamefont {Yabana}}, \bibinfo {author}
  {\bibfnamefont {T.}~\bibnamefont {Nakatsukasa}}, \ and\ \bibinfo {author}
  {\bibfnamefont {G.~F.}\ \bibnamefont {Bertsch}},\ }\bibfield  {title}
  {\enquote {\bibinfo {title} {First-principles electron dynamics simulation
  for optical breakdown of dielectrics under an intense laser field},}\
  }\href@noop {} {\bibfield  {journal} {\bibinfo  {journal} {Phys.~Rev.~B.~}\
  }\textbf {\bibinfo {volume} {77}},\ \bibinfo {pages} {165104} (\bibinfo
  {year} {2008})}\BibitemShut {NoStop}%
\bibitem [{\citenamefont {Lee}\ \emph {et~al.}(2014)\citenamefont {Lee},
  \citenamefont {Kim}, \citenamefont {Sato}, \citenamefont {Otobe},
  \citenamefont {Shinohara}, \citenamefont {Yabana},\ and\ \citenamefont
  {Moon~Jeong}}]{lee+2014jap}%
  \BibitemOpen
  \bibfield  {author} {\bibinfo {author} {\bibfnamefont {K.~M.}\ \bibnamefont
  {Lee}}, \bibinfo {author} {\bibfnamefont {C.}~\bibnamefont {Kim}}, \bibinfo
  {author} {\bibfnamefont {S.}~\bibnamefont {Sato}}, \bibinfo {author}
  {\bibfnamefont {T.}~\bibnamefont {Otobe}}, \bibinfo {author} {\bibfnamefont
  {Y.}~\bibnamefont {Shinohara}}, \bibinfo {author} {\bibfnamefont
  {K.}~\bibnamefont {Yabana}}, \ and\ \bibinfo {author} {\bibfnamefont
  {T.}~\bibnamefont {Moon~Jeong}},\ }\bibfield  {title} {\enquote {\bibinfo
  {title} {First-principles simulation of the optical response of bulk and
  thin-film $\alpha$-quartz irradiated with an ultrashort intense laser
  pulse},}\ }\href {\doibase 10.1063/1.4864662} {\bibfield  {journal} {\bibinfo
   {journal} {J.~Appl.~Phys.~}\ }\textbf {\bibinfo {volume} {115}},\ \bibinfo
  {pages} {053519} (\bibinfo {year} {2014})}\BibitemShut {NoStop}%
\bibitem [{\citenamefont {Walkenhorst}\ \emph {et~al.}(2016)\citenamefont
  {Walkenhorst}, \citenamefont {De~Giovannini}, \citenamefont {Castro},\ and\
  \citenamefont {Rubio}}]{wa+2016epjb}%
  \BibitemOpen
  \bibfield  {author} {\bibinfo {author} {\bibfnamefont {J.}~\bibnamefont
  {Walkenhorst}}, \bibinfo {author} {\bibfnamefont {U.}~\bibnamefont
  {De~Giovannini}}, \bibinfo {author} {\bibfnamefont {A.}~\bibnamefont
  {Castro}}, \ and\ \bibinfo {author} {\bibfnamefont {A.}~\bibnamefont
  {Rubio}},\ }\bibfield  {title} {\enquote {\bibinfo {title} {Tailored
  pump-probe transient spectroscopy with time-dependent density-functional
  theory: controlling absorption spectra},}\ }\href {\doibase
  10.1140/epjb/e2016-70064-0} {\bibfield  {journal} {\bibinfo  {journal}
  {Eur.~Phys.~J.~B.~}\ }\textbf {\bibinfo {volume} {89}},\ \bibinfo {pages}
  {128} (\bibinfo {year} {2016})}\BibitemShut {NoStop}%
\bibitem [{\citenamefont {Wachter}\ \emph {et~al.}(2014)\citenamefont
  {Wachter}, \citenamefont {Lemell}, \citenamefont {Burgd\"orfer},
  \citenamefont {Sato}, \citenamefont {Tong},\ and\ \citenamefont
  {Yabana}}]{wach+2014prl}%
  \BibitemOpen
  \bibfield  {author} {\bibinfo {author} {\bibfnamefont {G.}~\bibnamefont
  {Wachter}}, \bibinfo {author} {\bibfnamefont {C.}~\bibnamefont {Lemell}},
  \bibinfo {author} {\bibfnamefont {J.}~\bibnamefont {Burgd\"orfer}}, \bibinfo
  {author} {\bibfnamefont {S.~A.}\ \bibnamefont {Sato}}, \bibinfo {author}
  {\bibfnamefont {X.-M.}\ \bibnamefont {Tong}}, \ and\ \bibinfo {author}
  {\bibfnamefont {K.}~\bibnamefont {Yabana}},\ }\bibfield  {title} {\enquote
  {\bibinfo {title} {Ab initio simulation of electrical currents induced by
  ultrafast laser excitation of dielectric materials},}\ }\href@noop {}
  {\bibfield  {journal} {\bibinfo  {journal} {Phys. Rev. Lett.}\ }\textbf
  {\bibinfo {volume} {113}},\ \bibinfo {pages} {087401} (\bibinfo {year}
  {2014})}\BibitemShut {NoStop}%
\bibitem [{\citenamefont {Jacobs}\ \emph {et~al.}(2020)\citenamefont {Jacobs},
  \citenamefont {Krumland}, \citenamefont {Valencia}, \citenamefont {Wang},
  \citenamefont {Rossi},\ and\ \citenamefont {Cocchi}}]{matheusPaper}%
  \BibitemOpen
  \bibfield  {author} {\bibinfo {author} {\bibfnamefont {M.}~\bibnamefont
  {Jacobs}}, \bibinfo {author} {\bibfnamefont {J.}~\bibnamefont {Krumland}},
  \bibinfo {author} {\bibfnamefont {A.~M.}\ \bibnamefont {Valencia}}, \bibinfo
  {author} {\bibfnamefont {H.}~\bibnamefont {Wang}}, \bibinfo {author}
  {\bibfnamefont {M.}~\bibnamefont {Rossi}}, \ and\ \bibinfo {author}
  {\bibfnamefont {C.}~\bibnamefont {Cocchi}},\ }\bibfield  {title} {\enquote
  {\bibinfo {title} {Ultrafast charge transfer and vibronic coupling in a
  laser-excited hybrid inorganic/organic interface},}\ }\href@noop {}
  {\bibfield  {journal} {\bibinfo  {journal} {Advances in Physics: X}\ }\textbf
  {\bibinfo {volume} {5}},\ \bibinfo {pages} {1749883} (\bibinfo {year}
  {2020})}\BibitemShut {NoStop}%
\bibitem [{\citenamefont {Marques}\ and\ \citenamefont
  {Gross}(2004)}]{marq+ARPC2004}%
  \BibitemOpen
  \bibfield  {author} {\bibinfo {author} {\bibfnamefont {M.}~\bibnamefont
  {Marques}}\ and\ \bibinfo {author} {\bibfnamefont {E.}~\bibnamefont
  {Gross}},\ }\bibfield  {title} {\enquote {\bibinfo {title} {Time-dependent
  density functional theory},}\ }\href@noop {} {\bibfield  {journal} {\bibinfo
  {journal} {Annu.~Rev.~Phys.~Chem.~}\ }\textbf {\bibinfo {volume} {55}},\
  \bibinfo {pages} {427--455} (\bibinfo {year} {2004})}\BibitemShut {NoStop}%
\bibitem [{\citenamefont {A.~L.~Marques}\ \emph {et~al.}(2012)\citenamefont
  {A.~L.~Marques}, \citenamefont {Maitra}, \citenamefont {Nogueira},
  \citenamefont {Gross},\ and\ \citenamefont {Rubio}}]{marq+2012}%
  \BibitemOpen
  \bibfield  {author} {\bibinfo {author} {\bibfnamefont {M.}~\bibnamefont
  {A.~L.~Marques}}, \bibinfo {author} {\bibfnamefont {N.}~\bibnamefont
  {Maitra}}, \bibinfo {author} {\bibfnamefont {F.}~\bibnamefont {Nogueira}},
  \bibinfo {author} {\bibfnamefont {E.}~\bibnamefont {Gross}}, \ and\ \bibinfo
  {author} {\bibfnamefont {A.}~\bibnamefont {Rubio}},\ }\href@noop {} {\emph
  {\bibinfo {title} {Fundamentals of Time-Dependent Density Functional
  Theory}}},\ Vol.\ \bibinfo {volume} {837}\ (\bibinfo {year} {2012})\ p.\
  \bibinfo {pages} {130}\BibitemShut {NoStop}%
\bibitem [{\citenamefont {Andrade}\ \emph {et~al.}(2009)\citenamefont
  {Andrade}, \citenamefont {Castro}, \citenamefont {Zueco}, \citenamefont
  {Alonso}, \citenamefont {Echenique}, \citenamefont {Falceto},\ and\
  \citenamefont {Rubio}}]{andr+2009jctc}%
  \BibitemOpen
  \bibfield  {author} {\bibinfo {author} {\bibfnamefont {X.}~\bibnamefont
  {Andrade}}, \bibinfo {author} {\bibfnamefont {A.}~\bibnamefont {Castro}},
  \bibinfo {author} {\bibfnamefont {D.}~\bibnamefont {Zueco}}, \bibinfo
  {author} {\bibfnamefont {J.~L.}\ \bibnamefont {Alonso}}, \bibinfo {author}
  {\bibfnamefont {P.}~\bibnamefont {Echenique}}, \bibinfo {author}
  {\bibfnamefont {F.}~\bibnamefont {Falceto}}, \ and\ \bibinfo {author}
  {\bibfnamefont {n.}~\bibnamefont {Rubio}},\ }\bibfield  {title} {\enquote
  {\bibinfo {title} {Modified ehrenfest formalism for efficient large-scale ab
  initio molecular dynamics},}\ }\href@noop {} {\bibfield  {journal} {\bibinfo
  {journal} {J.~Chem.~Theory~Comput.~}\ }\textbf {\bibinfo {volume} {5}},\
  \bibinfo {pages} {728--742} (\bibinfo {year} {2009})}\BibitemShut {NoStop}%
\bibitem [{\citenamefont {Rozzi}, \citenamefont {Troiani},\ and\ \citenamefont
  {Tavernelli}(2017)}]{rozz+jcpm2017}%
  \BibitemOpen
  \bibfield  {author} {\bibinfo {author} {\bibfnamefont {C.~A.}\ \bibnamefont
  {Rozzi}}, \bibinfo {author} {\bibfnamefont {F.}~\bibnamefont {Troiani}}, \
  and\ \bibinfo {author} {\bibfnamefont {I.}~\bibnamefont {Tavernelli}},\
  }\bibfield  {title} {\enquote {\bibinfo {title} {Quantum modeling of
  ultrafast photoinduced charge separation},}\ }\href@noop {} {\bibfield
  {journal} {\bibinfo  {journal} {J.~Phys.~Condens.~Matter.~}\ }\textbf
  {\bibinfo {volume} {30}},\ \bibinfo {pages} {013002} (\bibinfo {year}
  {2017})}\BibitemShut {NoStop}%
\bibitem [{\citenamefont {Gallmann}, \citenamefont {Cirelli},\ and\
  \citenamefont {Keller}(2012)}]{ga+arpc2012}%
  \BibitemOpen
  \bibfield  {author} {\bibinfo {author} {\bibfnamefont {L.}~\bibnamefont
  {Gallmann}}, \bibinfo {author} {\bibfnamefont {C.}~\bibnamefont {Cirelli}}, \
  and\ \bibinfo {author} {\bibfnamefont {U.}~\bibnamefont {Keller}},\
  }\bibfield  {title} {\enquote {\bibinfo {title} {Attosecond science: Recent
  highlights and future trends},}\ }\href@noop {} {\bibfield  {journal}
  {\bibinfo  {journal} {Annu.~Rev.~Phys.~Chem.~}\ }\textbf {\bibinfo {volume}
  {63}},\ \bibinfo {pages} {447--469} (\bibinfo {year} {2012})}\BibitemShut
  {NoStop}%
\bibitem [{\citenamefont {Pazourek}, \citenamefont {Nagele},\ and\
  \citenamefont {Burgd\"orfer}(2015)}]{pazo+rmp2015}%
  \BibitemOpen
  \bibfield  {author} {\bibinfo {author} {\bibfnamefont {R.}~\bibnamefont
  {Pazourek}}, \bibinfo {author} {\bibfnamefont {S.}~\bibnamefont {Nagele}}, \
  and\ \bibinfo {author} {\bibfnamefont {J.}~\bibnamefont {Burgd\"orfer}},\
  }\bibfield  {title} {\enquote {\bibinfo {title} {Attosecond chronoscopy of
  photoemission},}\ }\href@noop {} {\bibfield  {journal} {\bibinfo  {journal}
  {Rev.~Mod.~Phys.~}\ }\textbf {\bibinfo {volume} {87}},\ \bibinfo {pages}
  {765--802} (\bibinfo {year} {2015})}\BibitemShut {NoStop}%
\bibitem [{\citenamefont {Landsman}\ and\ \citenamefont
  {Keller}(2015)}]{land+pr2015}%
  \BibitemOpen
  \bibfield  {author} {\bibinfo {author} {\bibfnamefont {A.~S.}\ \bibnamefont
  {Landsman}}\ and\ \bibinfo {author} {\bibfnamefont {U.}~\bibnamefont
  {Keller}},\ }\bibfield  {title} {\enquote {\bibinfo {title} {Attosecond
  science and the tunnelling time problem},}\ }\href {\doibase
  https://doi.org/10.1016/j.physrep.2014.09.002} {\bibfield  {journal}
  {\bibinfo  {journal} {Physics~Reports}\ }\textbf {\bibinfo {volume} {547}},\
  \bibinfo {pages} {1 -- 24} (\bibinfo {year} {2015})},\ \bibinfo {note}
  {attosecond science and the tunneling time problem}\BibitemShut {NoStop}%
\bibitem [{\citenamefont {Sommer}\ \emph {et~al.}(2016)\citenamefont {Sommer},
  \citenamefont {Bothschafter}, \citenamefont {Sato}, \citenamefont {Jakubeit},
  \citenamefont {Latka}, \citenamefont {Razskazovskaya}, \citenamefont
  {Fattahi}, \citenamefont {Jobst}, \citenamefont {Schweinberger},
  \citenamefont {Shirvanyan}, \citenamefont {Yakovlev}, \citenamefont
  {Kienberger}, \citenamefont {Yabana}, \citenamefont {Karpowicz},
  \citenamefont {Schultze},\ and\ \citenamefont {Krausz}}]{somm+16nature}%
  \BibitemOpen
  \bibfield  {author} {\bibinfo {author} {\bibfnamefont {A.}~\bibnamefont
  {Sommer}}, \bibinfo {author} {\bibfnamefont {E.}~\bibnamefont
  {Bothschafter}}, \bibinfo {author} {\bibfnamefont {S.}~\bibnamefont {Sato}},
  \bibinfo {author} {\bibfnamefont {C.}~\bibnamefont {Jakubeit}}, \bibinfo
  {author} {\bibfnamefont {T.}~\bibnamefont {Latka}}, \bibinfo {author}
  {\bibfnamefont {O.}~\bibnamefont {Razskazovskaya}}, \bibinfo {author}
  {\bibfnamefont {H.}~\bibnamefont {Fattahi}}, \bibinfo {author} {\bibfnamefont
  {M.}~\bibnamefont {Jobst}}, \bibinfo {author} {\bibfnamefont
  {W.}~\bibnamefont {Schweinberger}}, \bibinfo {author} {\bibfnamefont
  {V.}~\bibnamefont {Shirvanyan}}, \bibinfo {author} {\bibfnamefont
  {V.}~\bibnamefont {Yakovlev}}, \bibinfo {author} {\bibfnamefont
  {R.}~\bibnamefont {Kienberger}}, \bibinfo {author} {\bibfnamefont
  {K.}~\bibnamefont {Yabana}}, \bibinfo {author} {\bibfnamefont
  {N.}~\bibnamefont {Karpowicz}}, \bibinfo {author} {\bibfnamefont
  {M.}~\bibnamefont {Schultze}}, \ and\ \bibinfo {author} {\bibfnamefont
  {F.}~\bibnamefont {Krausz}},\ }\bibfield  {title} {\enquote {\bibinfo {title}
  {Attosecond nonlinear polarization and light--matter energy transfer in
  solids},}\ }\href@noop {} {\bibfield  {journal} {\bibinfo  {journal}
  {Nature}\ }\textbf {\bibinfo {volume} {534}},\ \bibinfo {pages} {86--90}
  (\bibinfo {year} {2016})}\BibitemShut {NoStop}%
\bibitem [{\citenamefont {De~Sio}\ \emph {et~al.}(2016)\citenamefont {De~Sio},
  \citenamefont {Troiani}, \citenamefont {Maiuri}, \citenamefont {R{\'e}hault},
  \citenamefont {Sommer}, \citenamefont {Lim}, \citenamefont {Huelga},
  \citenamefont {Plenio}, \citenamefont {Rozzi}, \citenamefont {Cerullo},
  \citenamefont {Molinari},\ and\ \citenamefont {Lienau}}]{desi+16natcom}%
  \BibitemOpen
  \bibfield  {author} {\bibinfo {author} {\bibfnamefont {A.}~\bibnamefont
  {De~Sio}}, \bibinfo {author} {\bibfnamefont {F.}~\bibnamefont {Troiani}},
  \bibinfo {author} {\bibfnamefont {M.}~\bibnamefont {Maiuri}}, \bibinfo
  {author} {\bibfnamefont {J.}~\bibnamefont {R{\'e}hault}}, \bibinfo {author}
  {\bibfnamefont {E.}~\bibnamefont {Sommer}}, \bibinfo {author} {\bibfnamefont
  {J.}~\bibnamefont {Lim}}, \bibinfo {author} {\bibfnamefont {S.~F.}\
  \bibnamefont {Huelga}}, \bibinfo {author} {\bibfnamefont {M.~B.}\
  \bibnamefont {Plenio}}, \bibinfo {author} {\bibfnamefont {C.~A.}\
  \bibnamefont {Rozzi}}, \bibinfo {author} {\bibfnamefont {G.}~\bibnamefont
  {Cerullo}}, \bibinfo {author} {\bibfnamefont {E.}~\bibnamefont {Molinari}}, \
  and\ \bibinfo {author} {\bibfnamefont {C.}~\bibnamefont {Lienau}},\
  }\bibfield  {title} {\enquote {\bibinfo {title} {Tracking the coherent
  generation of polaron pairs in conjugated polymers},}\ }\href@noop {}
  {\bibfield  {journal} {\bibinfo  {journal} {Nature Commun.}\ }\textbf
  {\bibinfo {volume} {7}},\ \bibinfo {pages} {13742} (\bibinfo {year}
  {2016})}\BibitemShut {NoStop}%
\bibitem [{\citenamefont {Schlaepfer}\ \emph {et~al.}(2018)\citenamefont
  {Schlaepfer}, \citenamefont {Lucchini}, \citenamefont {Sato}, \citenamefont
  {Volkov}, \citenamefont {Kasmi}, \citenamefont {Hartmann}, \citenamefont
  {Rubio}, \citenamefont {Gallmann},\ and\ \citenamefont
  {Keller}}]{schl+18natp}%
  \BibitemOpen
  \bibfield  {author} {\bibinfo {author} {\bibfnamefont {F.}~\bibnamefont
  {Schlaepfer}}, \bibinfo {author} {\bibfnamefont {M.}~\bibnamefont
  {Lucchini}}, \bibinfo {author} {\bibfnamefont {S.~A.}\ \bibnamefont {Sato}},
  \bibinfo {author} {\bibfnamefont {M.}~\bibnamefont {Volkov}}, \bibinfo
  {author} {\bibfnamefont {L.}~\bibnamefont {Kasmi}}, \bibinfo {author}
  {\bibfnamefont {N.}~\bibnamefont {Hartmann}}, \bibinfo {author}
  {\bibfnamefont {A.}~\bibnamefont {Rubio}}, \bibinfo {author} {\bibfnamefont
  {L.}~\bibnamefont {Gallmann}}, \ and\ \bibinfo {author} {\bibfnamefont
  {U.}~\bibnamefont {Keller}},\ }\bibfield  {title} {\enquote {\bibinfo {title}
  {Attosecond optical-field-enhanced carrier injection into the gaas conduction
  band},}\ }\href@noop {} {\bibfield  {journal} {\bibinfo  {journal} {Nature
  Phys.}\ }\textbf {\bibinfo {volume} {14}},\ \bibinfo {pages} {560--564}
  (\bibinfo {year} {2018})}\BibitemShut {NoStop}%
\bibitem [{\citenamefont {Buades}\ \emph {et~al.}(2018)\citenamefont {Buades},
  \citenamefont {Pic{\'o}n}, \citenamefont {Le{\'o}n}, \citenamefont {Di~Palo},
  \citenamefont {Cousin}, \citenamefont {Cocchi}, \citenamefont {Pellegrin},
  \citenamefont {Martin}, \citenamefont {Ma{\~n}as-Valero}, \citenamefont
  {Coronado}, \citenamefont {Danz}, \citenamefont {Draxl}, \citenamefont
  {Uemoto}, \citenamefont {Yabana}, \citenamefont {Schultze}, \citenamefont
  {Wall},\ and\ \citenamefont {Biegert}}]{buad+18}%
  \BibitemOpen
  \bibfield  {author} {\bibinfo {author} {\bibfnamefont {B.}~\bibnamefont
  {Buades}}, \bibinfo {author} {\bibfnamefont {A.}~\bibnamefont {Pic{\'o}n}},
  \bibinfo {author} {\bibfnamefont {I.}~\bibnamefont {Le{\'o}n}}, \bibinfo
  {author} {\bibfnamefont {N.}~\bibnamefont {Di~Palo}}, \bibinfo {author}
  {\bibfnamefont {S.~L.}\ \bibnamefont {Cousin}}, \bibinfo {author}
  {\bibfnamefont {C.}~\bibnamefont {Cocchi}}, \bibinfo {author} {\bibfnamefont
  {E.}~\bibnamefont {Pellegrin}}, \bibinfo {author} {\bibfnamefont {J.~H.}\
  \bibnamefont {Martin}}, \bibinfo {author} {\bibfnamefont {S.}~\bibnamefont
  {Ma{\~n}as-Valero}}, \bibinfo {author} {\bibfnamefont {E.}~\bibnamefont
  {Coronado}}, \bibinfo {author} {\bibfnamefont {T.}~\bibnamefont {Danz}},
  \bibinfo {author} {\bibfnamefont {C.}~\bibnamefont {Draxl}}, \bibinfo
  {author} {\bibfnamefont {M.}~\bibnamefont {Uemoto}}, \bibinfo {author}
  {\bibfnamefont {K.}~\bibnamefont {Yabana}}, \bibinfo {author} {\bibfnamefont
  {M.}~\bibnamefont {Schultze}}, \bibinfo {author} {\bibfnamefont
  {S.}~\bibnamefont {Wall}}, \ and\ \bibinfo {author} {\bibfnamefont
  {J.}~\bibnamefont {Biegert}},\ }\bibfield  {title} {\enquote {\bibinfo
  {title} {Attosecond-resolved petahertz carrier motion in semi-metallic
  tis2},}\ }\href@noop {} {\bibfield  {journal} {\bibinfo  {journal} {arXiv
  preprint arXiv:1808.06493}\ } (\bibinfo {year} {2018})}\BibitemShut {NoStop}%
\bibitem [{\citenamefont {Zangwill}\ and\ \citenamefont
  {Soven}(1980)}]{zang-sove80pra}%
  \BibitemOpen
  \bibfield  {author} {\bibinfo {author} {\bibfnamefont {A.}~\bibnamefont
  {Zangwill}}\ and\ \bibinfo {author} {\bibfnamefont {P.}~\bibnamefont
  {Soven}},\ }\bibfield  {title} {\enquote {\bibinfo {title}
  {Density-functional approach to local-field effects in finite systems:
  Photoabsorbtion in the rare gases},}\ }\href@noop {} {\bibfield  {journal}
  {\bibinfo  {journal} {Phys.~Rev.~A}\ }\textbf {\bibinfo {volume} {21}},\
  \bibinfo {pages} {1561--1572} (\bibinfo {year} {1980})}\BibitemShut {NoStop}%
\bibitem [{\citenamefont {Ekardt}(1985)}]{ekar1984prb}%
  \BibitemOpen
  \bibfield  {author} {\bibinfo {author} {\bibfnamefont {W.}~\bibnamefont
  {Ekardt}},\ }\bibfield  {title} {\enquote {\bibinfo {title} {Size-dependent
  photoabsorption and photoemission of small metal particles},}\ }\href@noop {}
  {\bibfield  {journal} {\bibinfo  {journal} {Phys.~Rev.~B}\ }\textbf {\bibinfo
  {volume} {31}},\ \bibinfo {pages} {6360--6370} (\bibinfo {year}
  {1985})}\BibitemShut {NoStop}%
\bibitem [{\citenamefont {Ekardt}(1984)}]{ekar1984prl}%
  \BibitemOpen
  \bibfield  {author} {\bibinfo {author} {\bibfnamefont {W.}~\bibnamefont
  {Ekardt}},\ }\bibfield  {title} {\enquote {\bibinfo {title} {Dynamical
  polarizability of small metal particles: Self-consistent spherical jellium
  background model},}\ }\href@noop {} {\bibfield  {journal} {\bibinfo
  {journal} {Phys.~Rev.~Lett.~}\ }\textbf {\bibinfo {volume} {52}},\ \bibinfo
  {pages} {1925--1928} (\bibinfo {year} {1984})}\BibitemShut {NoStop}%
\bibitem [{\citenamefont {Perdew}\ and\ \citenamefont
  {Wang}(1992)}]{perd-wang92prb}%
  \BibitemOpen
  \bibfield  {author} {\bibinfo {author} {\bibfnamefont {J.~P.}\ \bibnamefont
  {Perdew}}\ and\ \bibinfo {author} {\bibfnamefont {Y.}~\bibnamefont {Wang}},\
  }\bibfield  {title} {\enquote {\bibinfo {title} {Accurate and simple analytic
  representation of the electron-gas correlation energy},}\ }\href@noop {}
  {\bibfield  {journal} {\bibinfo  {journal} {Phys.~Rev.~B.~}\ }\textbf
  {\bibinfo {volume} {45}},\ \bibinfo {pages} {13244--13249} (\bibinfo {year}
  {1992})}\BibitemShut {NoStop}%
\bibitem [{\citenamefont {Runge}\ and\ \citenamefont
  {Gross}(1984)}]{rung+1984prl}%
  \BibitemOpen
  \bibfield  {author} {\bibinfo {author} {\bibfnamefont {E.}~\bibnamefont
  {Runge}}\ and\ \bibinfo {author} {\bibfnamefont {E.~K.~U.}\ \bibnamefont
  {Gross}},\ }\bibfield  {title} {\enquote {\bibinfo {title}
  {Density-functional theory for time-dependent systems},}\ }\href@noop {}
  {\bibfield  {journal} {\bibinfo  {journal} {Phys. Rev. Lett.}\ }\textbf
  {\bibinfo {volume} {52}},\ \bibinfo {pages} {997--1000} (\bibinfo {year}
  {1984})}\BibitemShut {NoStop}%
\bibitem [{\citenamefont {Maitra}, \citenamefont {Burke},\ and\ \citenamefont
  {Woodward}(2002)}]{mait+2002prl}%
  \BibitemOpen
  \bibfield  {author} {\bibinfo {author} {\bibfnamefont {N.~T.}\ \bibnamefont
  {Maitra}}, \bibinfo {author} {\bibfnamefont {K.}~\bibnamefont {Burke}}, \
  and\ \bibinfo {author} {\bibfnamefont {C.}~\bibnamefont {Woodward}},\
  }\bibfield  {title} {\enquote {\bibinfo {title} {Memory in time-dependent
  density functional theory},}\ }\href@noop {} {\bibfield  {journal} {\bibinfo
  {journal} {Phys.~Rev.~Lett.~}\ }\textbf {\bibinfo {volume} {89}},\ \bibinfo
  {pages} {023002} (\bibinfo {year} {2002})}\BibitemShut {NoStop}%
\bibitem [{\citenamefont {Gritsenko}\ and\ \citenamefont
  {Baerends}(2004)}]{grit+jcp2004}%
  \BibitemOpen
  \bibfield  {author} {\bibinfo {author} {\bibfnamefont {O.}~\bibnamefont
  {Gritsenko}}\ and\ \bibinfo {author} {\bibfnamefont {E.~J.}\ \bibnamefont
  {Baerends}},\ }\bibfield  {title} {\enquote {\bibinfo {title} {Asymptotic
  correction of the exchange-correlation kernel of time-dependent density
  functional theory for long-range charge-transfer excitations},}\ }\href
  {\doibase 10.1063/1.1759320} {\bibfield  {journal} {\bibinfo  {journal}
  {J.~Chem.~Phys.~}\ }\textbf {\bibinfo {volume} {121}},\ \bibinfo {pages}
  {655--660} (\bibinfo {year} {2004})}\BibitemShut {NoStop}%
\bibitem [{\citenamefont {Dreuw}\ and\ \citenamefont
  {Head-Gordon}(2004)}]{dreu+jacs2004}%
  \BibitemOpen
  \bibfield  {author} {\bibinfo {author} {\bibfnamefont {A.}~\bibnamefont
  {Dreuw}}\ and\ \bibinfo {author} {\bibfnamefont {M.}~\bibnamefont
  {Head-Gordon}},\ }\bibfield  {title} {\enquote {\bibinfo {title} {Failure of
  time-dependent density functional theory for long-range charge-transfer
  excited states: the zincbacteriochlorin- bacteriochlorin and
  bacteriochlorophyll- spheroidene complexes},}\ }\href@noop {} {\bibfield
  {journal} {\bibinfo  {journal} {J.~Am.~Chem.~Soc.~}\ }\textbf {\bibinfo
  {volume} {126}},\ \bibinfo {pages} {4007--4016} (\bibinfo {year}
  {2004})}\BibitemShut {NoStop}%
\bibitem [{\citenamefont {Autschbach}(2009)}]{autscpc2009}%
  \BibitemOpen
  \bibfield  {author} {\bibinfo {author} {\bibfnamefont {J.}~\bibnamefont
  {Autschbach}},\ }\bibfield  {title} {\enquote {\bibinfo {title}
  {Charge-transfer excitations and time-dependent density functional theory:
  Problems and some proposed solutions},}\ }\href@noop {} {\bibfield  {journal}
  {\bibinfo  {journal} {Chem.~Phys.~Chem.~}\ }\textbf {\bibinfo {volume}
  {10}},\ \bibinfo {pages} {1757--1760} (\bibinfo {year} {2009})}\BibitemShut
  {NoStop}%
\bibitem [{\citenamefont {Stein}, \citenamefont {Kronik},\ and\ \citenamefont
  {Baer}(2009)}]{stei+jaft2009}%
  \BibitemOpen
  \bibfield  {author} {\bibinfo {author} {\bibfnamefont {T.}~\bibnamefont
  {Stein}}, \bibinfo {author} {\bibfnamefont {L.}~\bibnamefont {Kronik}}, \
  and\ \bibinfo {author} {\bibfnamefont {R.}~\bibnamefont {Baer}},\ }\bibfield
  {title} {\enquote {\bibinfo {title} {Reliable prediction of charge transfer
  excitations in molecular complexes using time-dependent density functional
  theory},}\ }\href@noop {} {\bibfield  {journal} {\bibinfo  {journal}
  {J.~Am.~Chem.~Soc.~}\ }\textbf {\bibinfo {volume} {131}},\ \bibinfo {pages}
  {2818--2820} (\bibinfo {year} {2009})}\BibitemShut {NoStop}%
\bibitem [{\citenamefont {Rohrdanz}, \citenamefont {Martins},\ and\
  \citenamefont {Herbert}(2009)}]{rohr+jcp2009}%
  \BibitemOpen
  \bibfield  {author} {\bibinfo {author} {\bibfnamefont {M.~A.}\ \bibnamefont
  {Rohrdanz}}, \bibinfo {author} {\bibfnamefont {K.~M.}\ \bibnamefont
  {Martins}}, \ and\ \bibinfo {author} {\bibfnamefont {J.~M.}\ \bibnamefont
  {Herbert}},\ }\bibfield  {title} {\enquote {\bibinfo {title} {A
  long-range-corrected density functional that performs well for both
  ground-state properties and time-dependent density functional theory
  excitation energies, including charge-transfer excited states},}\ }\href@noop
  {} {\bibfield  {journal} {\bibinfo  {journal} {J.~Chem.~Phys.~}\ }\textbf
  {\bibinfo {volume} {130}},\ \bibinfo {pages} {054112} (\bibinfo {year}
  {2009})}\BibitemShut {NoStop}%
\bibitem [{\citenamefont {Maitra}(2017)}]{mait2017jpcm}%
  \BibitemOpen
  \bibfield  {author} {\bibinfo {author} {\bibfnamefont {N.~T.}\ \bibnamefont
  {Maitra}},\ }\bibfield  {title} {\enquote {\bibinfo {title} {Charge transfer
  in time-dependent density functional theory},}\ }\href@noop {} {\bibfield
  {journal} {\bibinfo  {journal} {J.~Phys.~Condens.~Matter.~}\ }\textbf
  {\bibinfo {volume} {29}},\ \bibinfo {pages} {423001} (\bibinfo {year}
  {2017})}\BibitemShut {NoStop}%
\bibitem [{\citenamefont {Campetella}\ \emph {et~al.}(2017)\citenamefont
  {Campetella}, \citenamefont {Maschietto}, \citenamefont {Frisch},
  \citenamefont {Scalmani}, \citenamefont {Ciofini},\ and\ \citenamefont
  {Adamo}}]{camp+jcc2017}%
  \BibitemOpen
  \bibfield  {author} {\bibinfo {author} {\bibfnamefont {M.}~\bibnamefont
  {Campetella}}, \bibinfo {author} {\bibfnamefont {F.}~\bibnamefont
  {Maschietto}}, \bibinfo {author} {\bibfnamefont {M.~J.}\ \bibnamefont
  {Frisch}}, \bibinfo {author} {\bibfnamefont {G.}~\bibnamefont {Scalmani}},
  \bibinfo {author} {\bibfnamefont {I.}~\bibnamefont {Ciofini}}, \ and\
  \bibinfo {author} {\bibfnamefont {C.}~\bibnamefont {Adamo}},\ }\bibfield
  {title} {\enquote {\bibinfo {title} {Charge transfer excitations in tddft: A
  ghost-hunter index},}\ }\href@noop {} {\bibfield  {journal} {\bibinfo
  {journal} {J.~Comput.~Chem.~}\ }\textbf {\bibinfo {volume} {38}},\ \bibinfo
  {pages} {2151--2156} (\bibinfo {year} {2017})}\BibitemShut {NoStop}%
\bibitem [{\citenamefont {Dreuw}, \citenamefont {Weisman},\ and\ \citenamefont
  {Head-Gordon}(2003)}]{dreu+2003jcp}%
  \BibitemOpen
  \bibfield  {author} {\bibinfo {author} {\bibfnamefont {A.}~\bibnamefont
  {Dreuw}}, \bibinfo {author} {\bibfnamefont {J.~L.}\ \bibnamefont {Weisman}},
  \ and\ \bibinfo {author} {\bibfnamefont {M.}~\bibnamefont {Head-Gordon}},\
  }\bibfield  {title} {\enquote {\bibinfo {title} {Long-range charge-transfer
  excited states in time-dependent density functional theory require non-local
  exchange},}\ }\href@noop {} {\bibfield  {journal} {\bibinfo  {journal}
  {J.~Chem.~Phys.~}\ }\textbf {\bibinfo {volume} {119}},\ \bibinfo {pages}
  {2943--2946} (\bibinfo {year} {2003})}\BibitemShut {NoStop}%
\bibitem [{\citenamefont {Tozer}(2003)}]{tozejcp2003}%
  \BibitemOpen
  \bibfield  {author} {\bibinfo {author} {\bibfnamefont {D.~J.}\ \bibnamefont
  {Tozer}},\ }\bibfield  {title} {\enquote {\bibinfo {title} {Relationship
  between long-range charge-transfer excitation energy error and integer
  discontinuity in Kohn-Sham theory},}\ }\href@noop {} {\bibfield  {journal}
  {\bibinfo  {journal} {J.~Chem.~Phys.~}\ }\textbf {\bibinfo {volume} {119}},\
  \bibinfo {pages} {12697--12699} (\bibinfo {year} {2003})}\BibitemShut
  {NoStop}%
\bibitem [{\citenamefont {Maitra}(2005)}]{maitjcp2005}%
  \BibitemOpen
  \bibfield  {author} {\bibinfo {author} {\bibfnamefont {N.~T.}\ \bibnamefont
  {Maitra}},\ }\bibfield  {title} {\enquote {\bibinfo {title} {Undoing static
  correlation: Long-range charge transfer in time-dependent density-functional
  theory},}\ }\href@noop {} {\bibfield  {journal} {\bibinfo  {journal}
  {J.~Chem.~Phys.~}\ }\textbf {\bibinfo {volume} {122}},\ \bibinfo {pages}
  {234104} (\bibinfo {year} {2005})}\BibitemShut {NoStop}%
\bibitem [{\citenamefont {Cocchi}\ and\ \citenamefont
  {Draxl}(2015)}]{cocc-drax15prb}%
  \BibitemOpen
  \bibfield  {author} {\bibinfo {author} {\bibfnamefont {C.}~\bibnamefont
  {Cocchi}}\ and\ \bibinfo {author} {\bibfnamefont {C.}~\bibnamefont {Draxl}},\
  }\bibfield  {title} {\enquote {\bibinfo {title} {Optical spectra from
  molecules to crystals: Insight from many-body perturbation theory},}\
  }\href@noop {} {\bibfield  {journal} {\bibinfo  {journal} {Phys.~Rev.~B}\
  }\textbf {\bibinfo {volume} {92}},\ \bibinfo {pages} {205126} (\bibinfo
  {year} {2015})}\BibitemShut {NoStop}%
\bibitem [{\citenamefont {Elliott}\ \emph {et~al.}(2011)\citenamefont
  {Elliott}, \citenamefont {Goldson}, \citenamefont {Canahui},\ and\
  \citenamefont {Maitra}}]{elli+chemphys2011}%
  \BibitemOpen
  \bibfield  {author} {\bibinfo {author} {\bibfnamefont {P.}~\bibnamefont
  {Elliott}}, \bibinfo {author} {\bibfnamefont {S.}~\bibnamefont {Goldson}},
  \bibinfo {author} {\bibfnamefont {C.}~\bibnamefont {Canahui}}, \ and\
  \bibinfo {author} {\bibfnamefont {N.~T.}\ \bibnamefont {Maitra}},\ }\bibfield
   {title} {\enquote {\bibinfo {title} {Perspectives on double-excitations in
  tddft},}\ }\href@noop {} {\bibfield  {journal} {\bibinfo  {journal}
  {Chem.~Phys.~}\ }\textbf {\bibinfo {volume} {391}},\ \bibinfo {pages} {110 --
  119} (\bibinfo {year} {2011})}\BibitemShut {NoStop}%
\bibitem [{\citenamefont {Fuks}\ \emph {et~al.}(2011)\citenamefont {Fuks},
  \citenamefont {Helbig}, \citenamefont {Tokatly},\ and\ \citenamefont
  {Rubio}}]{fuks+2011prb}%
  \BibitemOpen
  \bibfield  {author} {\bibinfo {author} {\bibfnamefont {J.~I.}\ \bibnamefont
  {Fuks}}, \bibinfo {author} {\bibfnamefont {N.}~\bibnamefont {Helbig}},
  \bibinfo {author} {\bibfnamefont {I.~V.}\ \bibnamefont {Tokatly}}, \ and\
  \bibinfo {author} {\bibfnamefont {A.}~\bibnamefont {Rubio}},\ }\bibfield
  {title} {\enquote {\bibinfo {title} {Nonlinear phenomena in time-dependent
  density-functional theory: What rabi oscillations can teach us},}\
  }\href@noop {} {\bibfield  {journal} {\bibinfo  {journal} {Phys.~Rev.~B}\
  }\textbf {\bibinfo {volume} {84}},\ \bibinfo {pages} {075107} (\bibinfo
  {year} {2011})}\BibitemShut {NoStop}%
\bibitem [{\citenamefont {Fuks}\ and\ \citenamefont
  {Maitra}(2014{\natexlab{a}})}]{fuks+pccp2014}%
  \BibitemOpen
  \bibfield  {author} {\bibinfo {author} {\bibfnamefont {J.~I.}\ \bibnamefont
  {Fuks}}\ and\ \bibinfo {author} {\bibfnamefont {N.~T.}\ \bibnamefont
  {Maitra}},\ }\bibfield  {title} {\enquote {\bibinfo {title} {Challenging
  adiabatic time-dependent density functional theory with a hubbard dimer: the
  case of time-resolved long-range charge transfer},}\ }\href@noop {}
  {\bibfield  {journal} {\bibinfo  {journal} {Phys.~Chem.~Chem.~Phys.~}\
  }\textbf {\bibinfo {volume} {16}},\ \bibinfo {pages} {14504--14513} (\bibinfo
  {year} {2014}{\natexlab{a}})}\BibitemShut {NoStop}%
\bibitem [{\citenamefont {Ullrich}\ and\ \citenamefont
  {Tokatly}(2006)}]{ullrich+prb2006}%
  \BibitemOpen
  \bibfield  {author} {\bibinfo {author} {\bibfnamefont {C.~A.}\ \bibnamefont
  {Ullrich}}\ and\ \bibinfo {author} {\bibfnamefont {I.~V.}\ \bibnamefont
  {Tokatly}},\ }\bibfield  {title} {\enquote {\bibinfo {title} {Nonadiabatic
  electron dynamics in time-dependent density-functional theory},}\ }\href@noop
  {} {\bibfield  {journal} {\bibinfo  {journal} {Phys.~Rev.~B}\ }\textbf
  {\bibinfo {volume} {73}},\ \bibinfo {pages} {235102} (\bibinfo {year}
  {2006})}\BibitemShut {NoStop}%
\bibitem [{\citenamefont {BostrÃ¶m}\ \emph {et~al.}(2018)\citenamefont
  {BostrÃ¶m}, \citenamefont {Mikkelsen}, \citenamefont {Verdozzi},
  \citenamefont {Perfetto},\ and\ \citenamefont {Stefanucci}}]{bostr2018nano}%
  \BibitemOpen
  \bibfield  {author} {\bibinfo {author} {\bibfnamefont {E.~V.}\ \bibnamefont
  {BostrÃ¶m}}, \bibinfo {author} {\bibfnamefont {A.}~\bibnamefont {Mikkelsen}},
  \bibinfo {author} {\bibfnamefont {C.}~\bibnamefont {Verdozzi}}, \bibinfo
  {author} {\bibfnamefont {E.}~\bibnamefont {Perfetto}}, \ and\ \bibinfo
  {author} {\bibfnamefont {G.}~\bibnamefont {Stefanucci}},\ }\bibfield  {title}
  {\enquote {\bibinfo {title} {Charge separation in donor-C60 complexes with
  real-time green functions: The importance of nonlocal correlations},}\
  }\href@noop {} {\bibfield  {journal} {\bibinfo  {journal} {Nano~Lett.~}\
  }\textbf {\bibinfo {volume} {18}},\ \bibinfo {pages} {785--792} (\bibinfo
  {year} {2018})}\BibitemShut {NoStop}%
\bibitem [{\citenamefont {Guandalini}\ \emph {et~al.}(2020)\citenamefont
  {Guandalini}, \citenamefont {Cocchi}, \citenamefont {Pittalis}, \citenamefont
  {Ruini},\ and\ \citenamefont {Rozzi}}]{albertosPaper}%
  \BibitemOpen
  \bibfield  {author} {\bibinfo {author} {\bibfnamefont {A.}~\bibnamefont
  {Guandalini}}, \bibinfo {author} {\bibfnamefont {C.}~\bibnamefont {Cocchi}},
  \bibinfo {author} {\bibfnamefont {S.}~\bibnamefont {Pittalis}}, \bibinfo
  {author} {\bibfnamefont {A.}~\bibnamefont {Ruini}}, \ and\ \bibinfo {author}
  {\bibfnamefont {C.}~\bibnamefont {Rozzi}},\ }\bibfield  {title} {\enquote
  {\bibinfo {title} {Nonlinear response of a quantum system to impulsive
  perturbations: A non-perturbative real-time approach},}\ }\href@noop {} {\
  (\bibinfo {year} {2020})}\BibitemShut {NoStop}%
\bibitem [{\citenamefont {Seidner}, \citenamefont {Stock},\ and\ \citenamefont
  {Domcke}(1995)}]{seid+1995jcp}%
  \BibitemOpen
  \bibfield  {author} {\bibinfo {author} {\bibfnamefont {L.}~\bibnamefont
  {Seidner}}, \bibinfo {author} {\bibfnamefont {G.}~\bibnamefont {Stock}}, \
  and\ \bibinfo {author} {\bibfnamefont {W.}~\bibnamefont {Domcke}},\
  }\bibfield  {title} {\enquote {\bibinfo {title} {Nonperturbative approach to
  femtosecond spectroscopy: General theory and application to multidimensional
  nonadiabatic photoisomerization processes},}\ }\href@noop {} {\bibfield
  {journal} {\bibinfo  {journal} {J.~Chem.~Phys.~}\ }\textbf {\bibinfo {volume}
  {103}},\ \bibinfo {pages} {3998--4011} (\bibinfo {year} {1995})}\BibitemShut
  {NoStop}%
\bibitem [{\citenamefont {Hamm}\ and\ \citenamefont
  {Zanni}(2011)}]{hamm_zanni_2011}%
  \BibitemOpen
  \bibfield  {author} {\bibinfo {author} {\bibfnamefont {P.}~\bibnamefont
  {Hamm}}\ and\ \bibinfo {author} {\bibfnamefont {M.}~\bibnamefont {Zanni}},\
  }\href {\doibase 10.1017/CBO9780511675935} {\emph {\bibinfo {title} {Concepts
  and Methods of 2D Infrared Spectroscopy}}}\ (\bibinfo  {publisher} {Cambridge
  University Press},\ \bibinfo {year} {2011})\BibitemShut {NoStop}%
\bibitem [{\citenamefont {Casida}(1996)}]{casi1996rda}%
  \BibitemOpen
  \bibfield  {author} {\bibinfo {author} {\bibfnamefont {M.}~\bibnamefont
  {Casida}},\ }\bibfield  {title} {\enquote {\bibinfo {title} {Theoretical and
  computational chemistry},}\ }\href@noop {} {\bibfield  {journal} {\bibinfo
  {journal} {Recent Developments and Applications in Modern Density Functional
  Theory}\ }\textbf {\bibinfo {volume} {4}} (\bibinfo {year}
  {1996})}\BibitemShut {NoStop}%
\bibitem [{\citenamefont {Casida}\ and\ \citenamefont
  {Huix-Rotllant}(2012)}]{casi+2012arpc}%
  \BibitemOpen
  \bibfield  {author} {\bibinfo {author} {\bibfnamefont {M.}~\bibnamefont
  {Casida}}\ and\ \bibinfo {author} {\bibfnamefont {M.}~\bibnamefont
  {Huix-Rotllant}},\ }\bibfield  {title} {\enquote {\bibinfo {title} {Progress
  in time-dependent density-functional theory},}\ }\href@noop {} {\bibfield
  {journal} {\bibinfo  {journal} {Annu.~Rev.~Phys.~Chem.~}\ }\textbf {\bibinfo
  {volume} {63}},\ \bibinfo {pages} {287--323} (\bibinfo {year}
  {2012})}\BibitemShut {NoStop}%
\bibitem [{\citenamefont {Hedin}(1965)}]{hedi+1965pr}%
  \BibitemOpen
  \bibfield  {author} {\bibinfo {author} {\bibfnamefont {L.}~\bibnamefont
  {Hedin}},\ }\bibfield  {title} {\enquote {\bibinfo {title} {New method for
  calculating the one-particle green's function with application to the
  electron-gas problem},}\ }\href@noop {} {\bibfield  {journal} {\bibinfo
  {journal} {Phys.~Rev.~}\ }\textbf {\bibinfo {volume} {139}},\ \bibinfo
  {pages} {A796--A823} (\bibinfo {year} {1965})}\BibitemShut {NoStop}%
\bibitem [{\citenamefont {Faber}\ \emph {et~al.}(2013)\citenamefont {Faber},
  \citenamefont {Boulanger}, \citenamefont {Duchemin}, \citenamefont
  {Attaccalite},\ and\ \citenamefont {Blase}}]{fabe+13jcp}%
  \BibitemOpen
  \bibfield  {author} {\bibinfo {author} {\bibfnamefont {C.}~\bibnamefont
  {Faber}}, \bibinfo {author} {\bibfnamefont {P.}~\bibnamefont {Boulanger}},
  \bibinfo {author} {\bibfnamefont {I.}~\bibnamefont {Duchemin}}, \bibinfo
  {author} {\bibfnamefont {C.}~\bibnamefont {Attaccalite}}, \ and\ \bibinfo
  {author} {\bibfnamefont {X.}~\bibnamefont {Blase}},\ }\bibfield  {title}
  {\enquote {\bibinfo {title} {Many-body green's function gw and bethe-salpeter
  study of the optical excitations in a paradigmatic model dipeptide},}\
  }\href@noop {} {\bibfield  {journal} {\bibinfo  {journal} {J.~Chem.~Phys.~}\
  }\textbf {\bibinfo {volume} {139}},\ \bibinfo {pages} {194308} (\bibinfo
  {year} {2013})}\BibitemShut {NoStop}%
\bibitem [{\citenamefont {Onida}, \citenamefont {Reining},\ and\ \citenamefont
  {Rubio}(2002)}]{onid+02rmp}%
  \BibitemOpen
  \bibfield  {author} {\bibinfo {author} {\bibfnamefont {G.}~\bibnamefont
  {Onida}}, \bibinfo {author} {\bibfnamefont {L.}~\bibnamefont {Reining}}, \
  and\ \bibinfo {author} {\bibfnamefont {A.}~\bibnamefont {Rubio}},\ }\bibfield
   {title} {\enquote {\bibinfo {title} {Electronic excitations:
  density-functional versus many-body Green's-function approaches},}\
  }\href@noop {} {\bibfield  {journal} {\bibinfo  {journal} {Rev.~Mod.~Phys.~}\
  }\textbf {\bibinfo {volume} {74}},\ \bibinfo {pages} {601} (\bibinfo {year}
  {2002})}\BibitemShut {NoStop}%
\bibitem [{\citenamefont {Bruneval}\ \emph {et~al.}(2016)\citenamefont
  {Bruneval}, \citenamefont {Rangel}, \citenamefont {Hamed}, \citenamefont
  {Shao}, \citenamefont {Yang},\ and\ \citenamefont {Neaton}}]{brun+16cpc}%
  \BibitemOpen
  \bibfield  {author} {\bibinfo {author} {\bibfnamefont {F.}~\bibnamefont
  {Bruneval}}, \bibinfo {author} {\bibfnamefont {T.}~\bibnamefont {Rangel}},
  \bibinfo {author} {\bibfnamefont {S.~M.}\ \bibnamefont {Hamed}}, \bibinfo
  {author} {\bibfnamefont {M.}~\bibnamefont {Shao}}, \bibinfo {author}
  {\bibfnamefont {C.}~\bibnamefont {Yang}}, \ and\ \bibinfo {author}
  {\bibfnamefont {J.~B.}\ \bibnamefont {Neaton}},\ }\bibfield  {title}
  {\enquote {\bibinfo {title} {molgw 1: Many-body perturbation theory software
  for atoms, molecules, and clusters},}\ }\href@noop {} {\bibfield  {journal}
  {\bibinfo  {journal} {Comput.~Phys.~Commun.~}\ }\textbf {\bibinfo {volume}
  {208}},\ \bibinfo {pages} {149 -- 161} (\bibinfo {year} {2016})}\BibitemShut
  {NoStop}%
\bibitem [{\citenamefont {Bruneval}, \citenamefont {Hamed},\ and\ \citenamefont
  {Neaton}(2015)}]{brun+jcp2015}%
  \BibitemOpen
  \bibfield  {author} {\bibinfo {author} {\bibfnamefont {F.}~\bibnamefont
  {Bruneval}}, \bibinfo {author} {\bibfnamefont {S.~M.}\ \bibnamefont {Hamed}},
  \ and\ \bibinfo {author} {\bibfnamefont {J.~B.}\ \bibnamefont {Neaton}},\
  }\bibfield  {title} {\enquote {\bibinfo {title} {A systematic benchmark of
  the ab initio bethe-salpeter equation approach for low-lying optical
  excitations of small organic molecules},}\ }\href@noop {} {\bibfield
  {journal} {\bibinfo  {journal} {J.~Chem.~Phys.~}\ }\textbf {\bibinfo {volume}
  {142}},\ \bibinfo {pages} {244101} (\bibinfo {year} {2015})}\BibitemShut
  {NoStop}%
\bibitem [{\citenamefont {Ullrich}(2012)}]{ullrich}%
  \BibitemOpen
  \bibfield  {author} {\bibinfo {author} {\bibfnamefont {C.~A.}\ \bibnamefont
  {Ullrich}},\ }\href {\doibase 10.1093/acprof:oso/9780199563029.001.0001}
  {\emph {\bibinfo {title} {Time-Dependent Density-Functional Theory: Concepts
  and Applications}}}\ (\bibinfo  {publisher} {Oxford University Press},\
  \bibinfo {year} {2012})\BibitemShut {NoStop}%
\bibitem [{\citenamefont {Marques}\ \emph {et~al.}(2003)\citenamefont
  {Marques}, \citenamefont {Castro}, \citenamefont {Bertsch},\ and\
  \citenamefont {Rubio}}]{marq+cpc2003}%
  \BibitemOpen
  \bibfield  {author} {\bibinfo {author} {\bibfnamefont {M.~A.}\ \bibnamefont
  {Marques}}, \bibinfo {author} {\bibfnamefont {A.}~\bibnamefont {Castro}},
  \bibinfo {author} {\bibfnamefont {G.~F.}\ \bibnamefont {Bertsch}}, \ and\
  \bibinfo {author} {\bibfnamefont {A.}~\bibnamefont {Rubio}},\ }\bibfield
  {title} {\enquote {\bibinfo {title} {octopus: a first-principles tool for
  excited electron-ion dynamics},}\ }\href@noop {} {\bibfield  {journal}
  {\bibinfo  {journal} {Comput.~Phys.~Commun.~}\ }\textbf {\bibinfo {volume}
  {151}},\ \bibinfo {pages} {60 -- 78} (\bibinfo {year} {2003})}\BibitemShut
  {NoStop}%
\bibitem [{\citenamefont {Castro}\ \emph {et~al.}(2006)\citenamefont {Castro},
  \citenamefont {Appel}, \citenamefont {Oliveira}, \citenamefont {Rozzi},
  \citenamefont {Andrade}, \citenamefont {Lorenzen}, \citenamefont {Marques},
  \citenamefont {Gross},\ and\ \citenamefont {Rubio}}]{cast+pssb2006}%
  \BibitemOpen
  \bibfield  {author} {\bibinfo {author} {\bibfnamefont {A.}~\bibnamefont
  {Castro}}, \bibinfo {author} {\bibfnamefont {H.}~\bibnamefont {Appel}},
  \bibinfo {author} {\bibfnamefont {M.}~\bibnamefont {Oliveira}}, \bibinfo
  {author} {\bibfnamefont {C.~A.}\ \bibnamefont {Rozzi}}, \bibinfo {author}
  {\bibfnamefont {X.}~\bibnamefont {Andrade}}, \bibinfo {author} {\bibfnamefont
  {F.}~\bibnamefont {Lorenzen}}, \bibinfo {author} {\bibfnamefont {M.~A.~L.}\
  \bibnamefont {Marques}}, \bibinfo {author} {\bibfnamefont {E.~K.~U.}\
  \bibnamefont {Gross}}, \ and\ \bibinfo {author} {\bibfnamefont
  {A.}~\bibnamefont {Rubio}},\ }\bibfield  {title} {\enquote {\bibinfo {title}
  {octopus: a tool for the application of time-dependent density functional
  theory},}\ }\href@noop {} {\bibfield  {journal} {\bibinfo  {journal}
  {Phys.~Status~Solidi~B}\ }\textbf {\bibinfo {volume} {243}},\ \bibinfo
  {pages} {2465--2488} (\bibinfo {year} {2006})}\BibitemShut {NoStop}%
\bibitem [{\citenamefont {Andrade}\ \emph {et~al.}(2015)\citenamefont
  {Andrade}, \citenamefont {Strubbe}, \citenamefont {De~Giovannini},
  \citenamefont {Larsen}, \citenamefont {Oliveira}, \citenamefont
  {Alberdi-Rodriguez}, \citenamefont {Varas}, \citenamefont {Theophilou},
  \citenamefont {Helbig}, \citenamefont {Verstraete}, \citenamefont {Stella},
  \citenamefont {Nogueira}, \citenamefont {Aspuru-Guzik}, \citenamefont
  {Castro}, \citenamefont {Marques},\ and\ \citenamefont
  {Rubio}}]{andr+pccp2015}%
  \BibitemOpen
  \bibfield  {author} {\bibinfo {author} {\bibfnamefont {X.}~\bibnamefont
  {Andrade}}, \bibinfo {author} {\bibfnamefont {D.}~\bibnamefont {Strubbe}},
  \bibinfo {author} {\bibfnamefont {U.}~\bibnamefont {De~Giovannini}}, \bibinfo
  {author} {\bibfnamefont {A.~H.}\ \bibnamefont {Larsen}}, \bibinfo {author}
  {\bibfnamefont {M.~J.~T.}\ \bibnamefont {Oliveira}}, \bibinfo {author}
  {\bibfnamefont {J.}~\bibnamefont {Alberdi-Rodriguez}}, \bibinfo {author}
  {\bibfnamefont {A.}~\bibnamefont {Varas}}, \bibinfo {author} {\bibfnamefont
  {I.}~\bibnamefont {Theophilou}}, \bibinfo {author} {\bibfnamefont
  {N.}~\bibnamefont {Helbig}}, \bibinfo {author} {\bibfnamefont {M.~J.}\
  \bibnamefont {Verstraete}}, \bibinfo {author} {\bibfnamefont
  {L.}~\bibnamefont {Stella}}, \bibinfo {author} {\bibfnamefont
  {F.}~\bibnamefont {Nogueira}}, \bibinfo {author} {\bibfnamefont
  {A.}~\bibnamefont {Aspuru-Guzik}}, \bibinfo {author} {\bibfnamefont
  {A.}~\bibnamefont {Castro}}, \bibinfo {author} {\bibfnamefont {M.~A.~L.}\
  \bibnamefont {Marques}}, \ and\ \bibinfo {author} {\bibfnamefont
  {A.}~\bibnamefont {Rubio}},\ }\bibfield  {title} {\enquote {\bibinfo {title}
  {Real-space grids and the octopus code as tools for the development of new
  simulation approaches for electronic systems},}\ }\href {\doibase
  10.1039/C5CP00351B} {\bibfield  {journal} {\bibinfo  {journal}
  {Phys.~Chem.~Chem.~Phys.~}\ }\textbf {\bibinfo {volume} {17}},\ \bibinfo
  {pages} {31371--31396} (\bibinfo {year} {2015})}\BibitemShut {NoStop}%
\bibitem [{\citenamefont {Tancogne-Dejean}\ \emph {et~al.}(2020)\citenamefont
  {Tancogne-Dejean}, \citenamefont {Oliveira}, \citenamefont {Andrade},
  \citenamefont {Appel}, \citenamefont {Borca}, \citenamefont {Le~Breton},
  \citenamefont {Buchholz}, \citenamefont {Castro}, \citenamefont {Corni},
  \citenamefont {Correa}, \citenamefont {De~Giovannini}, \citenamefont
  {Delgado}, \citenamefont {Eich}, \citenamefont {Flick}, \citenamefont {Gil},
  \citenamefont {Gomez}, \citenamefont {Helbig}, \citenamefont {HÃ¼bener},
  \citenamefont {JestÃ¤dt}, \citenamefont {Jornet-Somoza}, \citenamefont
  {Larsen}, \citenamefont {Lebedeva}, \citenamefont {LÃ¼ders}, \citenamefont
  {Marques}, \citenamefont {Ohlmann}, \citenamefont {Pipolo}, \citenamefont
  {Rampp}, \citenamefont {Rozzi}, \citenamefont {Strubbe}, \citenamefont
  {Sato}, \citenamefont {SchÃ¤fer}, \citenamefont {Theophilou}, \citenamefont
  {Welden},\ and\ \citenamefont {Rubio}}]{Octopus2020}%
  \BibitemOpen
  \bibfield  {author} {\bibinfo {author} {\bibfnamefont {N.}~\bibnamefont
  {Tancogne-Dejean}}, \bibinfo {author} {\bibfnamefont {M.~J.~T.}\ \bibnamefont
  {Oliveira}}, \bibinfo {author} {\bibfnamefont {X.}~\bibnamefont {Andrade}},
  \bibinfo {author} {\bibfnamefont {H.}~\bibnamefont {Appel}}, \bibinfo
  {author} {\bibfnamefont {C.~H.}\ \bibnamefont {Borca}}, \bibinfo {author}
  {\bibfnamefont {G.}~\bibnamefont {Le~Breton}}, \bibinfo {author}
  {\bibfnamefont {F.}~\bibnamefont {Buchholz}}, \bibinfo {author}
  {\bibfnamefont {A.}~\bibnamefont {Castro}}, \bibinfo {author} {\bibfnamefont
  {S.}~\bibnamefont {Corni}}, \bibinfo {author} {\bibfnamefont {A.~A.}\
  \bibnamefont {Correa}}, \bibinfo {author} {\bibfnamefont {U.}~\bibnamefont
  {De~Giovannini}}, \bibinfo {author} {\bibfnamefont {A.}~\bibnamefont
  {Delgado}}, \bibinfo {author} {\bibfnamefont {F.~G.}\ \bibnamefont {Eich}},
  \bibinfo {author} {\bibfnamefont {J.}~\bibnamefont {Flick}}, \bibinfo
  {author} {\bibfnamefont {G.}~\bibnamefont {Gil}}, \bibinfo {author}
  {\bibfnamefont {A.}~\bibnamefont {Gomez}}, \bibinfo {author} {\bibfnamefont
  {N.}~\bibnamefont {Helbig}}, \bibinfo {author} {\bibfnamefont
  {H.}~\bibnamefont {H\"ubener}}, \bibinfo {author} {\bibfnamefont
  {R.}~\bibnamefont {Jest\"adt}}, \bibinfo {author} {\bibfnamefont
  {J.}~\bibnamefont {Jornet-Somoza}}, \bibinfo {author} {\bibfnamefont {A.~H.}\
  \bibnamefont {Larsen}}, \bibinfo {author} {\bibfnamefont {I.~V.}\
  \bibnamefont {Lebedeva}}, \bibinfo {author} {\bibfnamefont {M.}~\bibnamefont
  {L\"uders}}, \bibinfo {author} {\bibfnamefont {M.~A.~L.}\ \bibnamefont
  {Marques}}, \bibinfo {author} {\bibfnamefont {S.~T.}\ \bibnamefont
  {Ohlmann}}, \bibinfo {author} {\bibfnamefont {S.}~\bibnamefont {Pipolo}},
  \bibinfo {author} {\bibfnamefont {M.}~\bibnamefont {Rampp}}, \bibinfo
  {author} {\bibfnamefont {C.~A.}\ \bibnamefont {Rozzi}}, \bibinfo {author}
  {\bibfnamefont {D.~A.}\ \bibnamefont {Strubbe}}, \bibinfo {author}
  {\bibfnamefont {S.~A.}\ \bibnamefont {Sato}}, \bibinfo {author}
  {\bibfnamefont {C.}~\bibnamefont {Sch\"afer}}, \bibinfo {author}
  {\bibfnamefont {I.}~\bibnamefont {Theophilou}}, \bibinfo {author}
  {\bibfnamefont {A.}~\bibnamefont {Welden}}, \ and\ \bibinfo {author}
  {\bibfnamefont {A.}~\bibnamefont {Rubio}},\ }\bibfield  {title} {\enquote
  {\bibinfo {title} {Octopus, a computational framework for exploring
  light-driven phenomena and quantum dynamics in extended and finite
  systems},}\ }\href {\doibase 10.1063/1.5142502} {\bibfield  {journal}
  {\bibinfo  {journal} {J.~Chem.~Phys.~}\ }\textbf {\bibinfo {volume} {152}},\
  \bibinfo {pages} {124119} (\bibinfo {year} {2020})}\BibitemShut {NoStop}%
\bibitem [{\citenamefont {Troullier}\ and\ \citenamefont
  {Martins}(1991)}]{trou-mart91prb}%
  \BibitemOpen
  \bibfield  {author} {\bibinfo {author} {\bibfnamefont {N.}~\bibnamefont
  {Troullier}}\ and\ \bibinfo {author} {\bibfnamefont {J.~L.}\ \bibnamefont
  {Martins}},\ }\bibfield  {title} {\enquote {\bibinfo {title} {Efficient
  pseudopotentials for plane-wave calculations},}\ }\href@noop {} {\bibfield
  {journal} {\bibinfo  {journal} {Phys.~Rev.~B}\ }\textbf {\bibinfo {volume}
  {43}},\ \bibinfo {pages} {1993--2006} (\bibinfo {year} {1991})}\BibitemShut
  {NoStop}%
\bibitem [{\citenamefont {Perdew}\ and\ \citenamefont
  {Zunger}(1981)}]{perd-zunger81prb}%
  \BibitemOpen
  \bibfield  {author} {\bibinfo {author} {\bibfnamefont {J.~P.}\ \bibnamefont
  {Perdew}}\ and\ \bibinfo {author} {\bibfnamefont {A.}~\bibnamefont
  {Zunger}},\ }\bibfield  {title} {\enquote {\bibinfo {title} {Self-interaction
  correction to density-functional approximations for many-electron systems},}\
  }\href@noop {} {\bibfield  {journal} {\bibinfo  {journal} {Phys.~Rev.~B}\
  }\textbf {\bibinfo {volume} {23}},\ \bibinfo {pages} {5048--5079} (\bibinfo
  {year} {1981})}\BibitemShut {NoStop}%
\bibitem [{\citenamefont {Castro}, \citenamefont {Marques},\ and\ \citenamefont
  {Rubio}(2004)}]{cast+jcp2004}%
  \BibitemOpen
  \bibfield  {author} {\bibinfo {author} {\bibfnamefont {A.}~\bibnamefont
  {Castro}}, \bibinfo {author} {\bibfnamefont {M.~A.~L.}\ \bibnamefont
  {Marques}}, \ and\ \bibinfo {author} {\bibfnamefont {A.}~\bibnamefont
  {Rubio}},\ }\bibfield  {title} {\enquote {\bibinfo {title} {Propagators for
  the time-dependent Kohn-Sham equations},}\ }\href@noop {} {\bibfield
  {journal} {\bibinfo  {journal} {J.~Chem.~Phys.~}\ }\textbf {\bibinfo {volume}
  {121}},\ \bibinfo {pages} {3425--3433} (\bibinfo {year} {2004})}\BibitemShut
  {NoStop}%
\bibitem [{\citenamefont {Bitzek}\ \emph {et~al.}(2006)\citenamefont {Bitzek},
  \citenamefont {Koskinen}, \citenamefont {G\"ahler}, \citenamefont {Moseler},\
  and\ \citenamefont {Gumbsch}}]{bitz+prl2006}%
  \BibitemOpen
  \bibfield  {author} {\bibinfo {author} {\bibfnamefont {E.}~\bibnamefont
  {Bitzek}}, \bibinfo {author} {\bibfnamefont {P.}~\bibnamefont {Koskinen}},
  \bibinfo {author} {\bibfnamefont {F.}~\bibnamefont {G\"ahler}}, \bibinfo
  {author} {\bibfnamefont {M.}~\bibnamefont {Moseler}}, \ and\ \bibinfo
  {author} {\bibfnamefont {P.}~\bibnamefont {Gumbsch}},\ }\bibfield  {title}
  {\enquote {\bibinfo {title} {Structural relaxation made simple},}\
  }\href@noop {} {\bibfield  {journal} {\bibinfo  {journal}
  {Phys.~Rev.~Lett.~}\ }\textbf {\bibinfo {volume} {97}},\ \bibinfo {pages}
  {170201} (\bibinfo {year} {2006})}\BibitemShut {NoStop}%
\bibitem [{\citenamefont {Bruneval}(2012)}]{brun12jcp}%
  \BibitemOpen
  \bibfield  {author} {\bibinfo {author} {\bibfnamefont {F.}~\bibnamefont
  {Bruneval}},\ }\bibfield  {title} {\enquote {\bibinfo {title} {Ionization
  energy of atoms obtained from gw self-energy or from random phase
  approximation total energies},}\ }\href@noop {} {\bibfield  {journal}
  {\bibinfo  {journal} {J.~Chem.~Phys.~}\ }\textbf {\bibinfo {volume} {136}},\
  \bibinfo {pages} {194107} (\bibinfo {year} {2012})}\BibitemShut {NoStop}%
\bibitem [{\citenamefont {Weigend}, \citenamefont {K\"{o}hn},\ and\
  \citenamefont {H\"{a}ttig}(2002)}]{weig+02}%
  \BibitemOpen
  \bibfield  {author} {\bibinfo {author} {\bibfnamefont {F.}~\bibnamefont
  {Weigend}}, \bibinfo {author} {\bibfnamefont {A.}~\bibnamefont {K\"{o}hn}}, \
  and\ \bibinfo {author} {\bibfnamefont {C.}~\bibnamefont {H\"{a}ttig}},\
  }\bibfield  {title} {\enquote {\bibinfo {title} {Efficient use of the
  correlation consistent basis sets in resolution of the identity mp2
  calculations},}\ }\href@noop {} {\bibfield  {journal} {\bibinfo  {journal}
  {J.~Chem.~Phys.~}\ }\textbf {\bibinfo {volume} {116}},\ \bibinfo {pages}
  {3175--3183} (\bibinfo {year} {2002})}\BibitemShut {NoStop}%
\bibitem [{\citenamefont {Yanai}, \citenamefont {Tew},\ and\ \citenamefont
  {Handy}(2004)}]{yana+2004cpl}%
  \BibitemOpen
  \bibfield  {author} {\bibinfo {author} {\bibfnamefont {T.}~\bibnamefont
  {Yanai}}, \bibinfo {author} {\bibfnamefont {D.}~\bibnamefont {Tew}}, \ and\
  \bibinfo {author} {\bibfnamefont {N.}~\bibnamefont {Handy}},\ }\bibfield
  {title} {\enquote {\bibinfo {title} {A new hybrid exchange-correlation
  functional using the coulomb-attenuating method (cam-b3lyp)},}\ }\href@noop
  {} {\bibfield  {journal} {\bibinfo  {journal} {Chem.~Phys.~Lett.~}\ }\textbf
  {\bibinfo {volume} {393}},\ \bibinfo {pages} {51--57} (\bibinfo {year}
  {2004})}\BibitemShut {NoStop}%
\bibitem [{\citenamefont {Bruneval}\ and\ \citenamefont
  {Marques}(2013)}]{brun+jctc2013}%
  \BibitemOpen
  \bibfield  {author} {\bibinfo {author} {\bibfnamefont {F.}~\bibnamefont
  {Bruneval}}\ and\ \bibinfo {author} {\bibfnamefont {M.~A.~L.}\ \bibnamefont
  {Marques}},\ }\bibfield  {title} {\enquote {\bibinfo {title} {Benchmarking
  the starting points of the gw approximation for molecules},}\ }\href@noop {}
  {\bibfield  {journal} {\bibinfo  {journal} {J.~Chem.~Theory~Comput.~}\
  }\textbf {\bibinfo {volume} {9}},\ \bibinfo {pages} {324--329} (\bibinfo
  {year} {2013})}\BibitemShut {NoStop}%
\bibitem [{\citenamefont {Platt}, \citenamefont {Klevens},\ and\ \citenamefont
  {Price}(1949)}]{ethSpectrum}%
  \BibitemOpen
  \bibfield  {author} {\bibinfo {author} {\bibfnamefont {J.~R.}\ \bibnamefont
  {Platt}}, \bibinfo {author} {\bibfnamefont {H.~B.}\ \bibnamefont {Klevens}},
  \ and\ \bibinfo {author} {\bibfnamefont {W.~C.}\ \bibnamefont {Price}},\
  }\bibfield  {title} {\enquote {\bibinfo {title} {Absorption intensities of
  ethylenes and acetylenes in the vacuum ultraviolet},}\ }\href@noop {}
  {\bibfield  {journal} {\bibinfo  {journal} {J.~Chem.~Phys.~}\ }\textbf
  {\bibinfo {volume} {17}},\ \bibinfo {pages} {466--469} (\bibinfo {year}
  {1949})}\BibitemShut {NoStop}%
\bibitem [{\citenamefont {Dreuw}\ and\ \citenamefont
  {Head-Gordon}(2005)}]{dreu+cr2005}%
  \BibitemOpen
  \bibfield  {author} {\bibinfo {author} {\bibfnamefont {A.}~\bibnamefont
  {Dreuw}}\ and\ \bibinfo {author} {\bibfnamefont {M.}~\bibnamefont
  {Head-Gordon}},\ }\bibfield  {title} {\enquote {\bibinfo {title}
  {Single-reference ab initio methods for the calculation of excited states of
  large molecules},}\ }\href@noop {} {\bibfield  {journal} {\bibinfo  {journal}
  {Chem.~Rev.~}\ }\textbf {\bibinfo {volume} {105}},\ \bibinfo {pages}
  {4009--4037} (\bibinfo {year} {2005})}\BibitemShut {NoStop}%
\bibitem [{\citenamefont {Legrand}, \citenamefont {Suraud},\ and\ \citenamefont
  {Reinhard}(2002)}]{Legrand2002}%
  \BibitemOpen
  \bibfield  {author} {\bibinfo {author} {\bibfnamefont {C.}~\bibnamefont
  {Legrand}}, \bibinfo {author} {\bibfnamefont {E.}~\bibnamefont {Suraud}}, \
  and\ \bibinfo {author} {\bibfnamefont {P.-G.}\ \bibnamefont {Reinhard}},\
  }\bibfield  {title} {\enquote {\bibinfo {title} {Comparison of
  self-interaction-corrections for metal clusters},}\ }\href@noop {} {\bibfield
   {journal} {\bibinfo  {journal} {J.~Phys.~B}\ }\textbf {\bibinfo {volume}
  {35}},\ \bibinfo {pages} {1115--1128} (\bibinfo {year} {2002})}\BibitemShut
  {NoStop}%
\bibitem [{\citenamefont {Becker}\ \emph {et~al.}(1996)\citenamefont {Becker},
  \citenamefont {Seixas~de Melo}, \citenamefont {MaÃ§anita},\ and\
  \citenamefont {Elisei}}]{thioSpectrum}%
  \BibitemOpen
  \bibfield  {author} {\bibinfo {author} {\bibfnamefont {R.~S.}\ \bibnamefont
  {Becker}}, \bibinfo {author} {\bibfnamefont {J.}~\bibnamefont {Seixas~de
  Melo}}, \bibinfo {author} {\bibfnamefont {A.~L.}\ \bibnamefont {MaÃ§anita}},
  \ and\ \bibinfo {author} {\bibfnamefont {F.}~\bibnamefont {Elisei}},\
  }\bibfield  {title} {\enquote {\bibinfo {title} {Comprehensive evaluation of
  the absorption, photophysical, energy transfer, structural, and theoretical
  properties of $\alpha$-oligothiophenes with one to seven rings},}\
  }\href@noop {} {\bibfield  {journal} {\bibinfo  {journal} {J.~Phys.~Chem.~}\
  }\textbf {\bibinfo {volume} {100}},\ \bibinfo {pages} {18683--18695}
  (\bibinfo {year} {1996})}\BibitemShut {NoStop}%
\bibitem [{\citenamefont {Koch}\ and\ \citenamefont
  {Otto}(1972)}]{benzeneSpectrum}%
  \BibitemOpen
  \bibfield  {author} {\bibinfo {author} {\bibfnamefont {E.}~\bibnamefont
  {Koch}}\ and\ \bibinfo {author} {\bibfnamefont {A.}~\bibnamefont {Otto}},\
  }\bibfield  {title} {\enquote {\bibinfo {title} {Optical absorption of
  benzene vapour for photon energies from 6 ev to 35 ev},}\ }\href {\doibase
  https://doi.org/10.1016/0009-2614(72)90011-5} {\bibfield  {journal} {\bibinfo
   {journal} {Chem.~Phys.~Lett.~}\ }\textbf {\bibinfo {volume} {12}},\ \bibinfo
  {pages} {476 -- 480} (\bibinfo {year} {1972})}\BibitemShut {NoStop}%
\bibitem [{\citenamefont {Fuks}\ \emph {et~al.}(2015)\citenamefont {Fuks},
  \citenamefont {Luo}, \citenamefont {Sandoval},\ and\ \citenamefont
  {Maitra}}]{fuks+prl2015}%
  \BibitemOpen
  \bibfield  {author} {\bibinfo {author} {\bibfnamefont {J.~I.}\ \bibnamefont
  {Fuks}}, \bibinfo {author} {\bibfnamefont {K.}~\bibnamefont {Luo}}, \bibinfo
  {author} {\bibfnamefont {E.~D.}\ \bibnamefont {Sandoval}}, \ and\ \bibinfo
  {author} {\bibfnamefont {N.~T.}\ \bibnamefont {Maitra}},\ }\bibfield  {title}
  {\enquote {\bibinfo {title} {Time-resolved spectroscopy in time-dependent
  density functional theory: An exact condition},}\ }\href@noop {} {\bibfield
  {journal} {\bibinfo  {journal} {Phys.~Rev.~Lett.~}\ }\textbf {\bibinfo
  {volume} {114}},\ \bibinfo {pages} {183002} (\bibinfo {year}
  {2015})}\BibitemShut {NoStop}%
\bibitem [{\citenamefont {Luo}, \citenamefont {Fuks},\ and\ \citenamefont
  {Maitra}(2016)}]{luo2016}%
  \BibitemOpen
  \bibfield  {author} {\bibinfo {author} {\bibfnamefont {K.}~\bibnamefont
  {Luo}}, \bibinfo {author} {\bibfnamefont {J.~I.}\ \bibnamefont {Fuks}}, \
  and\ \bibinfo {author} {\bibfnamefont {N.~T.}\ \bibnamefont {Maitra}},\
  }\bibfield  {title} {\enquote {\bibinfo {title} {Studies of spuriously
  shifting resonances in time-dependent density functional theory},}\ }\href
  {\doibase 10.1063/1.4955447} {\bibfield  {journal} {\bibinfo  {journal}
  {J.~Chem.~Phys.~}\ }\textbf {\bibinfo {volume} {145}},\ \bibinfo {pages}
  {044101} (\bibinfo {year} {2016})}\BibitemShut {NoStop}%
\bibitem [{\citenamefont {Provorse}, \citenamefont {Habenicht},\ and\
  \citenamefont {Isborn}(2015)}]{prov+2015jctc}%
  \BibitemOpen
  \bibfield  {author} {\bibinfo {author} {\bibfnamefont {M.~R.}\ \bibnamefont
  {Provorse}}, \bibinfo {author} {\bibfnamefont {B.~F.}\ \bibnamefont
  {Habenicht}}, \ and\ \bibinfo {author} {\bibfnamefont {C.~M.}\ \bibnamefont
  {Isborn}},\ }\bibfield  {title} {\enquote {\bibinfo {title} {Peak-shifting in
  real-time time-dependent density functional theory},}\ }\href@noop {}
  {\bibfield  {journal} {\bibinfo  {journal} {J.~Chem.~Theory~Comput.~}\
  }\textbf {\bibinfo {volume} {11}},\ \bibinfo {pages} {4791--4802} (\bibinfo
  {year} {2015})}\BibitemShut {NoStop}%
\bibitem [{\citenamefont {Fuks}\ \emph {et~al.}(2013)\citenamefont {Fuks},
  \citenamefont {Elliott}, \citenamefont {Rubio},\ and\ \citenamefont
  {Maitra}}]{fuks2013}%
  \BibitemOpen
  \bibfield  {author} {\bibinfo {author} {\bibfnamefont {J.~I.}\ \bibnamefont
  {Fuks}}, \bibinfo {author} {\bibfnamefont {P.}~\bibnamefont {Elliott}},
  \bibinfo {author} {\bibfnamefont {A.}~\bibnamefont {Rubio}}, \ and\ \bibinfo
  {author} {\bibfnamefont {N.~T.}\ \bibnamefont {Maitra}},\ }\bibfield  {title}
  {\enquote {\bibinfo {title} {Dynamics of charge-transfer processes with
  time-dependent density functional theory},}\ }\href@noop {} {\bibfield
  {journal} {\bibinfo  {journal} {J.~Phys.~Chem.~Lett.}\ }\textbf {\bibinfo
  {volume} {4}},\ \bibinfo {pages} {735--739} (\bibinfo {year}
  {2013})}\BibitemShut {NoStop}%
\bibitem [{\citenamefont {Fuks}\ and\ \citenamefont
  {Maitra}(2014{\natexlab{b}})}]{fuks2014pra}%
  \BibitemOpen
  \bibfield  {author} {\bibinfo {author} {\bibfnamefont {J.~I.}\ \bibnamefont
  {Fuks}}\ and\ \bibinfo {author} {\bibfnamefont {N.~T.}\ \bibnamefont
  {Maitra}},\ }\bibfield  {title} {\enquote {\bibinfo {title} {Charge transfer
  in time-dependent density-functional theory: Insights from the asymmetric
  hubbard dimer},}\ }\href@noop {} {\bibfield  {journal} {\bibinfo  {journal}
  {Phys.~Rev.~A}\ }\textbf {\bibinfo {volume} {89}},\ \bibinfo {pages} {062502}
  (\bibinfo {year} {2014}{\natexlab{b}})}\BibitemShut {NoStop}%
\bibitem [{\citenamefont {Raghunathan}\ and\ \citenamefont
  {Nest}(2011)}]{ragh2011}%
  \BibitemOpen
  \bibfield  {author} {\bibinfo {author} {\bibfnamefont {S.}~\bibnamefont
  {Raghunathan}}\ and\ \bibinfo {author} {\bibfnamefont {M.}~\bibnamefont
  {Nest}},\ }\bibfield  {title} {\enquote {\bibinfo {title} {Critical
  examination of explicitly time-dependent density functional theory for
  coherent control of dipole switching},}\ }\href@noop {} {\bibfield  {journal}
  {\bibinfo  {journal} {J.~Chem.~Theory~Comput.~}\ }\textbf {\bibinfo {volume}
  {7}},\ \bibinfo {pages} {2492--2497} (\bibinfo {year} {2011})}\BibitemShut
  {NoStop}%
\bibitem [{\citenamefont {Raghunathan}\ and\ \citenamefont
  {Nest}(2012)}]{ragh2012_2}%
  \BibitemOpen
  \bibfield  {author} {\bibinfo {author} {\bibfnamefont {S.}~\bibnamefont
  {Raghunathan}}\ and\ \bibinfo {author} {\bibfnamefont {M.}~\bibnamefont
  {Nest}},\ }\bibfield  {title} {\enquote {\bibinfo {title} {Coherent control
  and time-dependent density functional theory: Towards creation of wave
  packets by ultrashort laser pulses},}\ }\href@noop {} {\bibfield  {journal}
  {\bibinfo  {journal} {J.~Chem.~Phys.~}\ }\textbf {\bibinfo {volume} {136}},\
  \bibinfo {pages} {064104} (\bibinfo {year} {2012})}\BibitemShut {NoStop}%
\bibitem [{\citenamefont {Provorse}\ and\ \citenamefont
  {Isborn}(2016)}]{prov2016}%
  \BibitemOpen
  \bibfield  {author} {\bibinfo {author} {\bibfnamefont {M.~R.}\ \bibnamefont
  {Provorse}}\ and\ \bibinfo {author} {\bibfnamefont {C.~M.}\ \bibnamefont
  {Isborn}},\ }\bibfield  {title} {\enquote {\bibinfo {title} {Electron
  dynamics with real-time time-dependent density functional theory},}\ }\href
  {\doibase 10.1002/qua.25096} {\bibfield  {journal} {\bibinfo  {journal}
  {Int.~J.~Quantum~Chem.}\ }\textbf {\bibinfo {volume} {116}},\ \bibinfo
  {pages} {739--749} (\bibinfo {year} {2016})}\BibitemShut {NoStop}%
\bibitem [{\citenamefont {Elliott}\ \emph {et~al.}(2012)\citenamefont
  {Elliott}, \citenamefont {Fuks}, \citenamefont {Rubio},\ and\ \citenamefont
  {Maitra}}]{elli+2012prl}%
  \BibitemOpen
  \bibfield  {author} {\bibinfo {author} {\bibfnamefont {P.}~\bibnamefont
  {Elliott}}, \bibinfo {author} {\bibfnamefont {J.~I.}\ \bibnamefont {Fuks}},
  \bibinfo {author} {\bibfnamefont {A.}~\bibnamefont {Rubio}}, \ and\ \bibinfo
  {author} {\bibfnamefont {N.~T.}\ \bibnamefont {Maitra}},\ }\bibfield  {title}
  {\enquote {\bibinfo {title} {Universal dynamical steps in the exact
  time-dependent exchange-correlation potential},}\ }\href@noop {} {\bibfield
  {journal} {\bibinfo  {journal} {Phys.~Rev.~Lett.~}\ }\textbf {\bibinfo
  {volume} {109}},\ \bibinfo {pages} {266404} (\bibinfo {year}
  {2012})}\BibitemShut {NoStop}%
\bibitem [{\citenamefont {Maitra}(2016)}]{mait2016}%
  \BibitemOpen
  \bibfield  {author} {\bibinfo {author} {\bibfnamefont {N.~T.}\ \bibnamefont
  {Maitra}},\ }\bibfield  {title} {\enquote {\bibinfo {title} {Perspective:
  Fundamental aspects of time-dependent density functional theory},}\
  }\href@noop {} {\bibfield  {journal} {\bibinfo  {journal} {J.~Chem.~Phys.~}\
  }\textbf {\bibinfo {volume} {144}},\ \bibinfo {pages} {220901} (\bibinfo
  {year} {2016})}\BibitemShut {NoStop}%
\bibitem [{\citenamefont {Perfetto}\ and\ \citenamefont
  {Stefanucci}(2015)}]{perf+2015pra}%
  \BibitemOpen
  \bibfield  {author} {\bibinfo {author} {\bibfnamefont {E.}~\bibnamefont
  {Perfetto}}\ and\ \bibinfo {author} {\bibfnamefont {G.}~\bibnamefont
  {Stefanucci}},\ }\bibfield  {title} {\enquote {\bibinfo {title} {Some exact
  properties of the nonequilibrium response function for transient
  photoabsorption},}\ }\href@noop {} {\bibfield  {journal} {\bibinfo  {journal}
  {Phys. Rev. A}\ }\textbf {\bibinfo {volume} {91}},\ \bibinfo {pages} {033416}
  (\bibinfo {year} {2015})}\BibitemShut {NoStop}%
\bibitem [{\citenamefont {Perdew}, \citenamefont {Burke},\ and\ \citenamefont
  {Ernzerhof}(1996)}]{pbe}%
  \BibitemOpen
  \bibfield  {author} {\bibinfo {author} {\bibfnamefont {J.~P.}\ \bibnamefont
  {Perdew}}, \bibinfo {author} {\bibfnamefont {K.}~\bibnamefont {Burke}}, \
  and\ \bibinfo {author} {\bibfnamefont {M.}~\bibnamefont {Ernzerhof}},\
  }\bibfield  {title} {\enquote {\bibinfo {title} {Generalized gradient
  approximation made simple},}\ }\href {\doibase 10.1103/PhysRevLett.77.3865}
  {\bibfield  {journal} {\bibinfo  {journal} {Phys. Rev. Lett.}\ }\textbf
  {\bibinfo {volume} {77}},\ \bibinfo {pages} {3865--3868} (\bibinfo {year}
  {1996})}\BibitemShut {NoStop}%
\end{thebibliography}

%

\end{document}